\documentclass[12pt,a4paper]{article}
\usepackage[T2A]{fontenc}
\usepackage[utf8]{inputenc}
\usepackage[english]{babel}
\usepackage{indentfirst}
\usepackage{fullpage}
\usepackage{url}
\usepackage{amsmath}
\usepackage{amssymb}
\usepackage{graphicx}
\usepackage{multirow}
\usepackage{array}
\usepackage{diagbox}
\usepackage{aas_macros}

\renewcommand{\vec}[1]{{\boldsymbol{#1}}}

\title{Tomo-V -- a New Tool for Doppler Tomography}
\author{P.V. Kaygorodov}

\begin{document}
\maketitle

\abstract{
In this paper, a new tool for Doppler tomography, Tomo-V (\url{https://tomo-v.inasan.ru}) that is developed based on the algebraic reconstruction technique (ART) has been presented. Previously, the ART method has not been widely used in tomography, as its direct implementation was computationally complex. The author has developed a fast version of this algorithm, which allowed it to be implemented within a web application that runs at acceptable speed in a browser on a personal computer. This method can be used to obtain sharp tomographic images from blurred profiles. Furthermore, the method has demonstrated excellent results in reconstructing images from noisy data, from a small number of profiles, and from profiles contaminated by absorption lines and emission from the expanding envelope. Tomo-V also includes tools for analyzing the resulting tomograms, allowing the position of accretion disks and Roche lobes to be displayed on the tomogram, as well as back-projecting the tomographic image onto flow elements in spatial coordinates. The paper is partially based on a report presented at the Modern Stellar Astronomy 2025 conference.}

\section{Introduction}

Interacting binary stars are often the only source of data on physical processes occurring in conditions unattainable on Earth: at high densities and temperatures, in the presence of strong gravitational and electromagnetic fields. For the majority of the most interesting binary systems, direct observation of individual flow elements is impossible, and modern observational tools can only measure integral quantities: light curves and spectra, to which all flow elements contribute simultaneously. Currently, Doppler tomography is the most advanced method for obtaining information about individual elements of the flow in a binary star based on variations in the spectral line profiles of binary stars.

To construct Doppler tomograms, methods similar to those used to construct tomographic images in other fields-medicine, flaw detection, etc.-are used. The input data for constructing Doppler tomograms are spectral line profiles. The peculiarity of these data, which distinguishes them from input data obtained, for example, during X-ray studies, is their heterogeneity (the state of the object can change during the observation process), noise, rather low resolution, and also the relatively small number of profiles that can be obtained for one object. To obtain tomograms from such data, it is necessary to use special approaches, including regularization methods, which allows  recovery of lost information from input data. The most commonly used method is the maximum entropy method (MEM)~\cite{1988MNRAS.235..269M}, which reconstructs images with the least possible amount of detail, which is important for interpretation. Also, a radio astronomical approach (CLEAN method)~\cite{1997ASPC..125..202A} is used to construct tomograms, which is well suited for identifying the boundaries of elements. These methods use a property of tomograms called the ``Central projection theorem'', which makes it possible to search for a solution in Fourier space by filling in regions missing from the initial data according to some algorithm.

In contrast, in the algebraic approach (ART)~\cite{GORDON1970471}, minimization methods are employed to find the solution directly (for Doppler tomography) in velocity space, without using the Fourier transform. Until now, this approach has not been used to obtain Doppler tomograms due probably to its high computational complexity: it requires minimization in a space with a dimension equal to the number of pixels in the desired image. For example, for an image with a resolution of $50\times 50$ pixels, it is needed to perform the minimization (like gradient descent) in a space with a dimension of 2500.

When developing Tomo-V, through certain optimizations, it was possible to achieve good performance when using the ART method, which made it possible to develop a program that runs directly in the browser and now it is available for public use at~\url{https://tomo-v.inasan.ru}. In this paper, a description of the used method, instructions for using the program, as well as the results of its testing on synthetic and observational data are provided.

\section{The ART Method}

The essence of the ART method consists of searching for a two-dimensional distribution of brightness in velocity space, such that the projections of this distribution that are taken for phases corresponding to the phases of the observed profiles fits best with the observed profiles. The standard deviation $\chi^2$ is used as a criterion of coincidence:
\begin{equation}
\chi^2=\sum_{p}\sum_{k=1}^{n_p}(I_{p,k}-I^*_{p,k})^2\,\,,\label{chi2}
\end{equation}
where $I_{p,k}$ is the intensity of $k$-th point of the observed profile $p$, while $I_{p,k}^*$ is the intensity of the corresponding point of the synthetic profile obtained by convolving the tomographic image.

In Tomo-V, each pixel in an image contributes to the brightness as a point source blurred as a Gaussian profile with a specified full width at half maximum (FWHM):
\begin{equation}
I^*_{p,k}=\sum_{i,j}^{n_x,n_y}dx\,dy\,I_{i,j}\exp\left(-\left(\dfrac{2\,\log 2}{W}\left(v_{p,k}-(\vec{v}_{i,j}\cdot\vec{l}_p)\right)\right)^2\right)\,\,,\label{weights}
\end{equation}
where $W$ is the FWHM value, $n_x,n_y$ are the tomogram resolution in pixels,  $dx$ and $dy$ are the pixel sizes, $v_{p,k}$ is the velocity at a point $k$ of profile $p$, $\vec{v}_{i,j}$ is the velocity at a point $i,j$ of the tomogram, $I_{i,j}$ is an intensity of the pixel and $\vec{l}_k$ is a
unit vector in velocity coordinates depending on the
phase of profile $p$:
$$
\vec{l}_p=-(\cos \phi_p,\sin{\phi_p})\,\,,
$$
where $\phi_p$ is the phase, at which the profile $p$ is taken. As $W$, the FWHM of the point spread function (PSF) added to the estimated FWHM of the source function associated with thermal line broadening can be taken. Due to the overlap of the Gaussian profiles of the points, a nonlinearity appears that prevents the solution from degenerating into a set of superimposed bands. To select the $W$ value, the following formula can be used:
$$
W\approx 0.24\sqrt{T}\,\,\text{(km/s)},
$$
where $T$ is the temperature of the radiating gas, K.

Before minimization begins, for each point of the image, its contribution to the brightness of each point of the observed spectra is calculated. To reduce memory consumption and speed up calculations, the contribution within the radius, where the amplitude of its Gaussian profile is greater than 2\% of the maximum, is taken into account for each pixel. After this, minimization is performed using the gradient descent method: the change in $\chi^2$ is determined (taking into account all profiles) by adding a small value to the brightness of each pixel, after which the entire image is updated: a ``step'' is taken along the calculated gradient by a certain value $\Delta$. When calculating the gradient, there is no need to recalculate completely all the terms in equation~\eqref{chi2}, since each pixel affects only a relatively small number of them. So that, from the initial $\chi^2$ value, the values that are affected by a given pixel are simply subtracted, after which the sum of similar terms, which are calculated with allowance for the change in this pixel by a small amount, is added.

The step size $\Delta$ is selected automatically. At the beginning, it is equal to some heuristic value and is adjusted during the minimization. For example, if the step resulted in an increase in $\chi^2$, value $\Delta$ is reduced by half and the step is repeated until the next step leads to a reduction in $\chi^2$ or $\Delta$ does decrease to a certain threshold value (in this case, the computing stops). If the step was successful (the value $\chi^2$ decreased), $\Delta$ increases by 10\%. As tests showed, this approach makes it possible to reduce quickly the $\Delta$ value to the optimal one and keep it close to the optimal one throughout the calculation.

Since each pixel in the image contributes as a Gaussian profile, the method effectively ``removes blur'': all details in the resulting solution will have sizes and shapes that match the real ones, and the image will not be ``convolved'' with the Gaussian profile of the source. This distinguishes the ART method from, for example, the maximum entropy method, which does not use information about line broadening and, accordingly, produces a blurred image. As tests showed, the method produces sufficiently clear images at FWHM of the source up to 10–20\% of the image size. At larger FWHM, the image may still be
somewhat blurry, but will still be sharper than the initial image convolved with the corresponding Gaussian profile.

Classically, when obtaining a Doppler tomogram, it is assumed that there is no absorption in the system. The presence of strong absorption in the line that is associated, for example, with the presence of an expanding envelope in a system in an active state makes it impossible to construct a Doppler tomogram using traditional methods. However, since the ART method uses minimization to fit the image to match best the observations, there is nothing preventing from introducing additional parameters into it for fitting. For example, it becomes entirely possible to select simultaneously an absorption profile applied on synthetic (adjusted) profiles in such a way as to obtain the best match to observations. In this case, in the observed profiles, we can see, for example, only a part of the S-waves formed by the flow elements, since the other part of them is in the zone with high absorption. However, with good coverage of the orbital period phases, it is possible to reconstruct the full image, since different elements are outside the absorption line at different times. Furthermore, even when located in an area, where absorption is strong, these elements influence the profile shape, which can be used for recovery. Besides absorption, profiles may also contain additional emission that does not change depending on the observation phase. This could be, for example, radiation from an expanding envelope, which, together with the absorption line, could contribute in the form of a P~Cyg-type profile.

To take into account the absorption line as well as the additional emission, equation~\eqref{chi2} was modified as follows:
\begin{equation}
\chi^2=\sum_{p}\sum_{k=1}^{n_p}(I_{p,k}-A(v_{p,k})\,I^*_{p,k}-E(v_{p,k}))^2\,\,,\label{chi2b}
\end{equation}
where $A(v_{p,k})=\exp(-\tau(v_{p,k}))$ is the absorption coefficient in the line for the velocity $v_{p,k}$ (the velocity at point $k$ of profile $p$) that is determined by the optical thickness of the envelope, $\tau$, in this part of the spectrum, and $E(v_{p,k})$ is the envelope emission in the same region. Thus, in Tomo-V, it becomes possible to select an absorption profile in such a way that the projections taken for the corresponding phases from the resulting image that are convolved with this profile best match the observed data.

When selecting absorption and additional emission profiles, additional smoothing is applied to both profiles. This allows for improved convergence at the initial stage when the system has a high chance of sliding into a local minimum by fitting the data primarily by selecting absorption and emission profiles rather than by selecting a tomogram image. When this happens, the program can usually eventually come to the correct solution, but the number of required steps increases many times over, and a lot of unnecessary detail appears on the absorption profile. In general, the fewer details there are in the absorption and additional radiation profiles, the more of them there can be in the tomogram, so it makes sense to look for the smoothest shape of these profiles. It is also likely more physical, since the expanding envelope is unlikely to be composed of many different flow elements and must give a relatively broad, near-Gaussian profile for both absorption and additional emission.

\section{Tomo-V Program Operation}

Tomo-V is a web application and available on the website, supported by Institute of Astronomy, Russian Academy of Sciences (INASAN) (\url{https://tomo-v.inasan.ru}). On the page, the user will see a screen divided into two parts: in the left column (initially empty), there is a list of saved images, an area for loading previously exported data (files with the .tmv extension, see below), and a button for creating a new image. When clicking this button, the user will be offered a choice: to create a test image (``Sample data'' for mastering the program, as well as for experimenting with the settings), or an image constructed from real data (``Real data'').

\begin{figure}[t]
	\begin{center}
		\includegraphics[width=16cm]{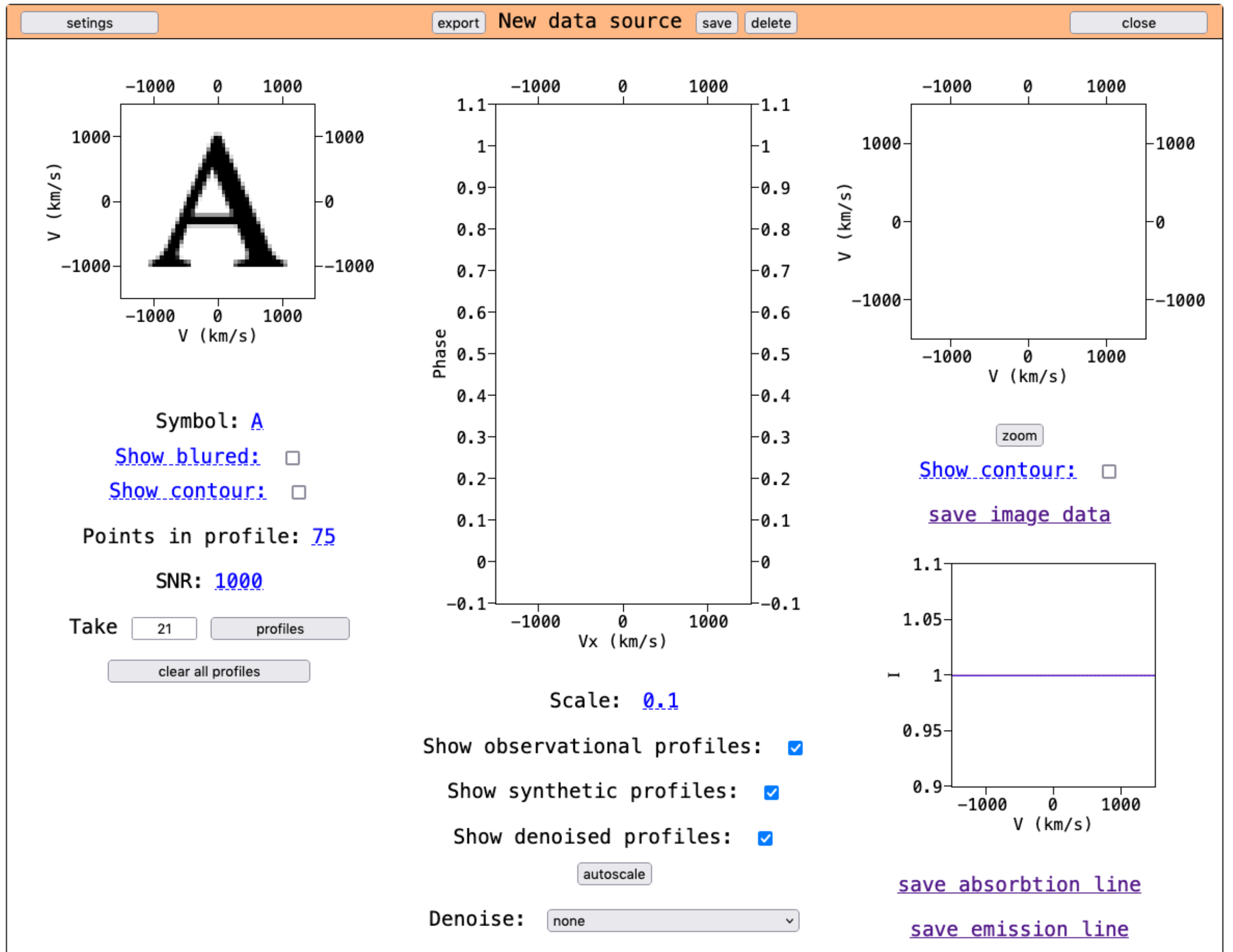}
	\end{center}
\caption{Working with synthetic data.}\label{SampleWin}
\end{figure}

If ``Sample data'' is selected, a window will open on the right side of the screen (see Fig.~\ref{SampleWin}), in the upper left part of which there will be an image of a test symbol (by default, the letter ``A''), a settings block are below, where one can change the symbol, enable display with blur (``Show blurred''), display the symbol outline (``Show contour''), change the number of points in synthetic profiles (``Points in profile''), the signal-tonoise ratio (``SNR''), as well as the ``Take X profiles'' buttons for generating the required number of profiles with uniform phase distribution and ``clear profiles'' to delete all profiles. By hovering the mouse cursor over the block with the symbol, one can select a phase and click on it to add a profile for the desired phase. In this way, one can test the operation of the method with an uneven distribution of profiles across phases. Next to this block is a graph for displaying profiles as a ``trailed spectra'' -- with a vertical shift depending on the phase. The same graph will show ``synthetic'' profiles (in blue) obtained from the reconstructed tomogram, as well as averaged profiles in green (see below). By comparing ``real'' (red) profiles with ``synthetic'' (blue) ones, the quality of tomogram reconstruction can be controlled.

The right part of the window shows the reconstructed image, a graph with the absorption line (blue) and additional emission (red), as well as elements related to the settings and control of the solution search process (not visible in Fig. 1). To deal with noisy source data, the ``Low SNR mode'' option can be enabled. The ``Noise treshhold'' parameter can be used to adjust the amount of clipping when ``Low SNR'' mode is enabled (see below). To fit (simultaneously with the tomogram) the absorption line and additional emission profile, it is necessary to enable the ``Fit absorption'' and ``Fit extra emission'' options, respectively. To reset the image, as well as the absorption/emission profiles to their initial state, the ``Reset solver'' button can be used; to reset only the absorption or emission lines, use the ``Reset absorption'' and ``Reset emission'' buttons, respectively. The ``Start reconstruction'' button is used to start the image construction process. The construction process can be stopped by clicking the ``Stop reconstruction'' button and continued using the ``Continue reconstruction'' button. The links ``save image data'', ``save absorption line'', and ``save emission line'' are used to save files with the results of image construction to disk. The files are in text format. For profiles, they consist of two columns, where the first column is the velocity in km/s, and the second is either the intensity (for the emission line) or the absorption coefficient (for the absorption line). The first two columns of the image file contain the velocity coordinates of the pixel center (also in km/s), and the third contains the intensity in arbitrary units. The window header also contains buttons for saving the state (the data is stored in the browser), exporting the state to a file (with a .tmv extension; the file can then be imported into another browser), deleting the saved state, and closing the window. Furthermore, in the upper left part of the window, there is a button for setting general settings: the speed range, FWHM of the PSF and resolution of the tomogram.

\begin{figure}[t]
	\begin{center}
		\includegraphics[width=16cm]{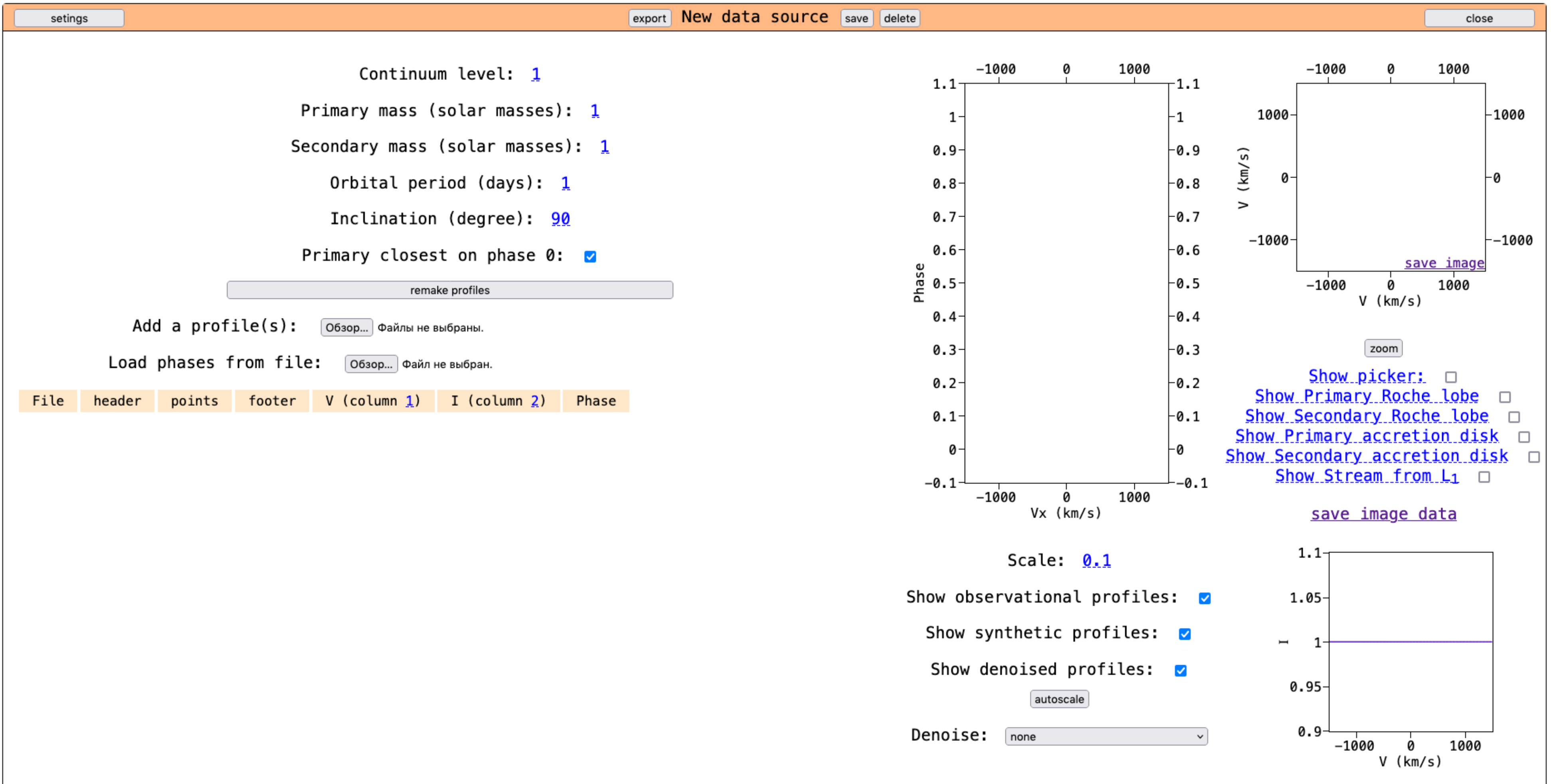}
	\end{center}
\caption{Working with real data.}\label{RealWin}
\end{figure}

If one selects ``Real data\dots'' when creating a new image, a window for constructing an image based on real data will open (see Fig. 2). In general, it is similar to the test one, but on the left side, there will be a button for selecting files, a table of loaded files, input fields for setting the continuum level (unity by default), and system parameters (orbital inclination, component masses, and orbital period). Uploaded files must be text files and contain two columns: velocity (in km/s) and intensity. By default, the velocity is taken from the first column, and the intensity is taken from the second column, but this can be changed, including after loading the files, by clicking on the corresponding table header. Intensities must be normalized to the continuum level, preferably so that the intensity of the continuum is taken as unity. Profiles may have different numbers of points, and the velocity sets of each profile do not have to match. For each downloaded file, it is necessary to specify the phase (in the right part of the table, then click the ``update'' button). Also, in the right part of the downloaded files table, opposite each file, there is a check box; by unchecking it, the corresponding profile can be removed from the calculations. Loaded profiles (the part of them that fits into the specified velocity range) are immediately shown on the graph to the right of the file table. A file specifying phases can also be uploaded (``Load phases from file'').

\section{Testing on Artificial Data}

\begin{figure}[ht]
  \begin{center}
	\begin{tabular}{ccc}
      \raisebox{-\totalheight}{\includegraphics[height=3cm]{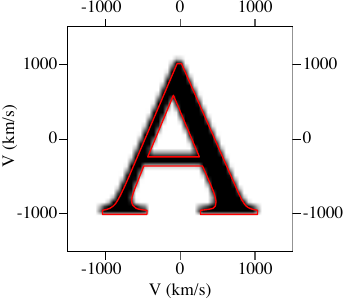}} & 
      \multirow{2}{*}{\raisebox{-\totalheight}{\includegraphics[height=6cm]{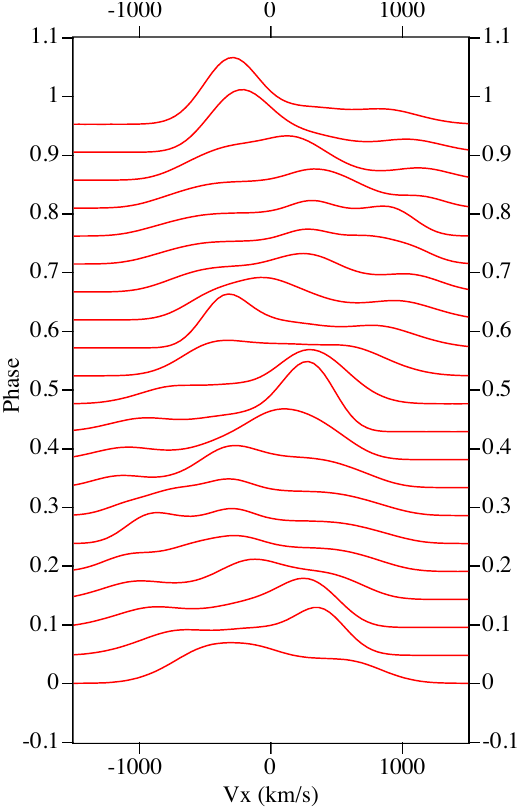}}} &
      \multirow{2}{*}{\raisebox{-\totalheight}{\includegraphics[height=6cm]{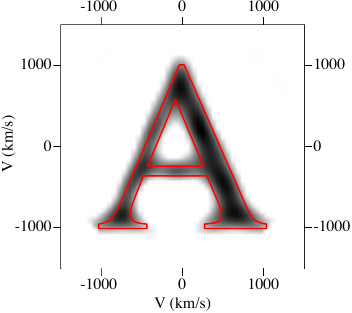}}}\\
      \raisebox{-\totalheight}{\includegraphics[height=3cm]{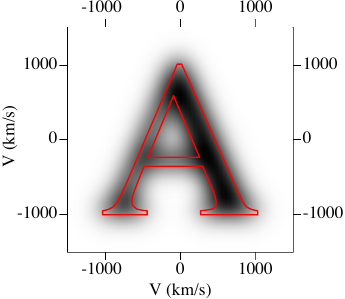}} & &\\
    \end{tabular}
  \end{center}
\caption{Initial image (left column, without blur and with blur), profiles (middle column) and reconstructed image (right column). FWHM = 100, SNR = 1000, $50\times 50$ pixels, 25 profiles, 75 points per profile, and 500 iterations.}\label{letterA1}
\end{figure}

The ``Sample data'' mode is designed for testing the recovery algorithm with various initial data. To begin with, we will show that the recovery operates in the ``ideal case'' in the absence of noise in the initial data (all tests can be repeated independently using the web application). In Fig.~\ref{letterA1}, the initial image and the profiles taken from it for phase 21 are shown. The FWHM of the blur was set to 100 km/s. As can be seen from the figure, almost complete coincidence of the red (``observed'') and blue (``model'') profiles is achieved. It is also clear that the algorithm has almost completely removed the blurring added to the initial profiles, restoring a sharp image.

\begin{figure}[ht]
  \begin{center}
	\begin{tabular}{ccc}
      \raisebox{-\totalheight}{\includegraphics[height=3cm]{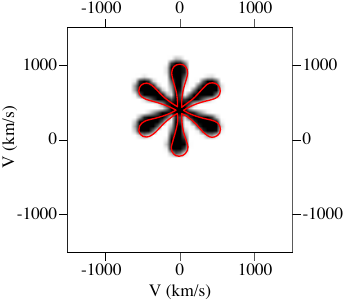}} & 
      \multirow{2}{*}{\raisebox{-\totalheight}{\includegraphics[height=6cm]{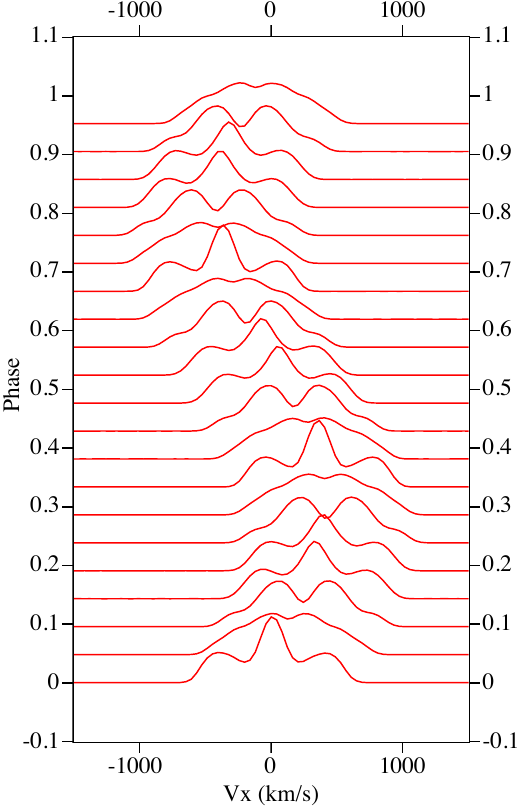}}} &
      \multirow{2}{*}{\raisebox{-\totalheight}{\includegraphics[height=6cm]{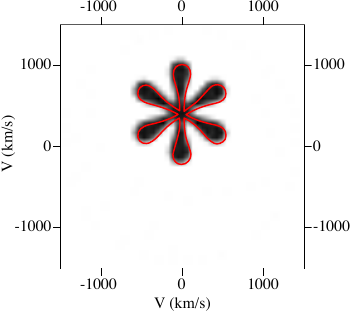}}}\\
      \raisebox{-\totalheight}{\includegraphics[height=3cm]{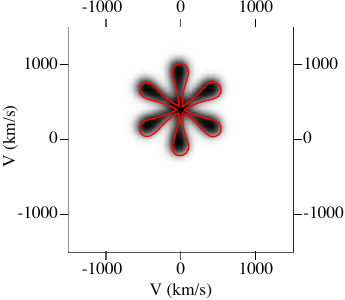}} & &\\
    \end{tabular}
  \end{center}
\caption{Similar to Fig.~\ref{letterA1} for the ``*'' symbol.}\label{star1}
\end{figure}

\begin{figure}[ht]
  \begin{center}
	\begin{tabular}{ccc}
      \raisebox{-\totalheight}{\includegraphics[height=3cm]{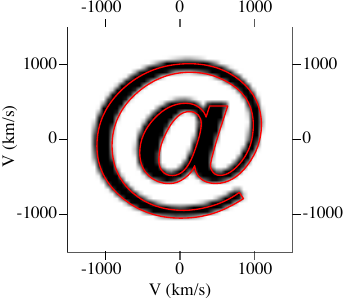}} & 
      \multirow{2}{*}{\raisebox{-\totalheight}{\includegraphics[height=6cm]{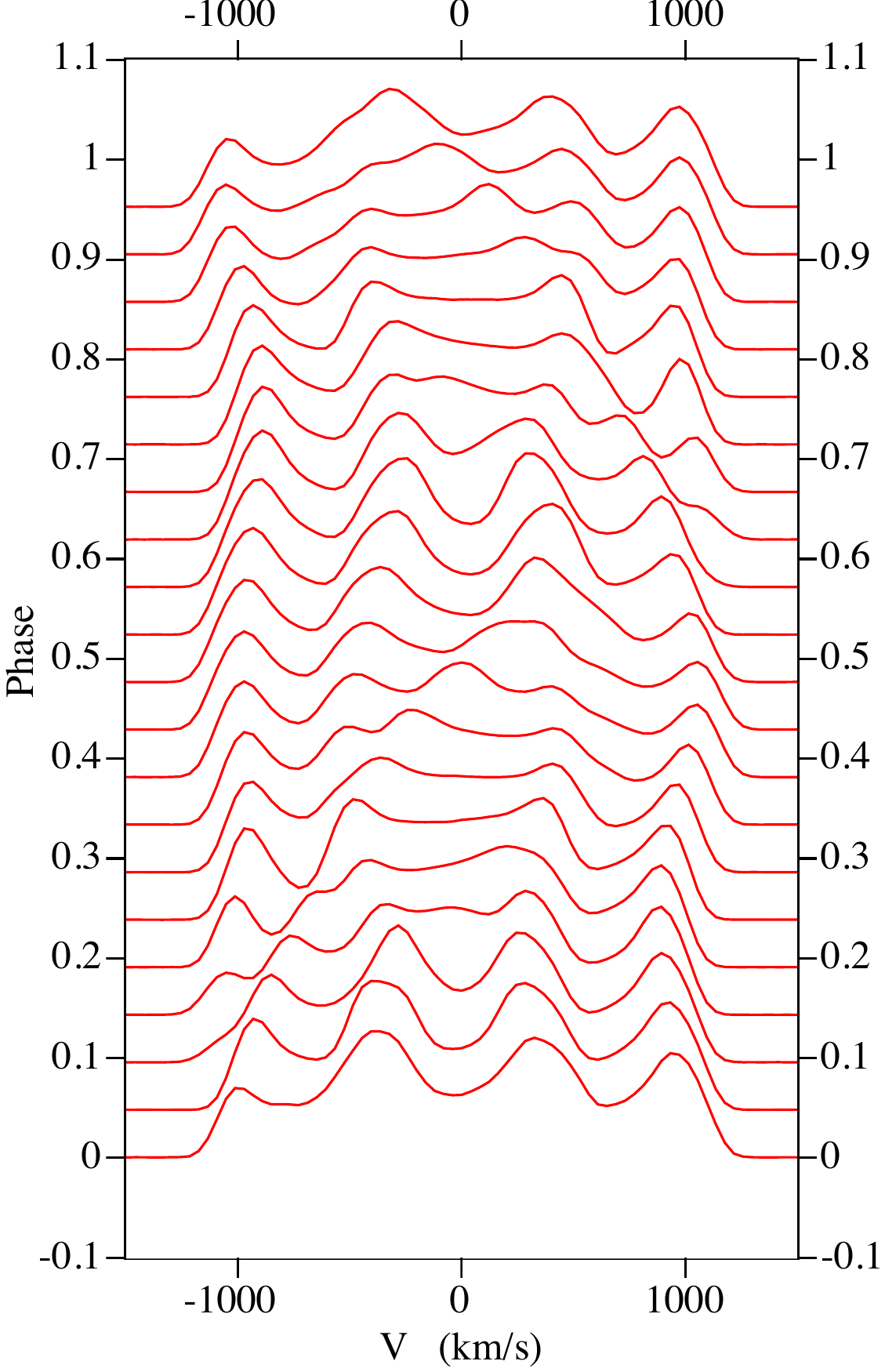}}} &
      \multirow{2}{*}{\raisebox{-\totalheight}{\includegraphics[height=6cm]{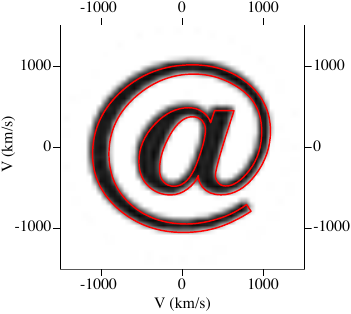}}}\\
      \raisebox{-\totalheight}{\includegraphics[height=3cm]{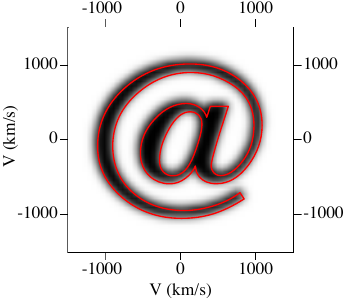}} & &\\
    \end{tabular}
  \end{center}
\caption{Similar to Fig.~\ref{letterA1} for the ``@'' symbol.}\label{dog1}
\end{figure}

In Figs.~\ref{star1} and~\ref{dog1}, a similar recovery is shown for the ``*'' and ``@'' symbols. As can be seen, the algorithm handled them as well. Particularly noteworthy is the complete absence of artifacts in the reconstructed images.

\subsection{Reconstruction By a Small Number of Profiles}

\begin{figure}[ht]
  \begin{center}
	\begin{tabular}{cccc}
      \raisebox{-\totalheight}{\includegraphics[width=3.5cm]{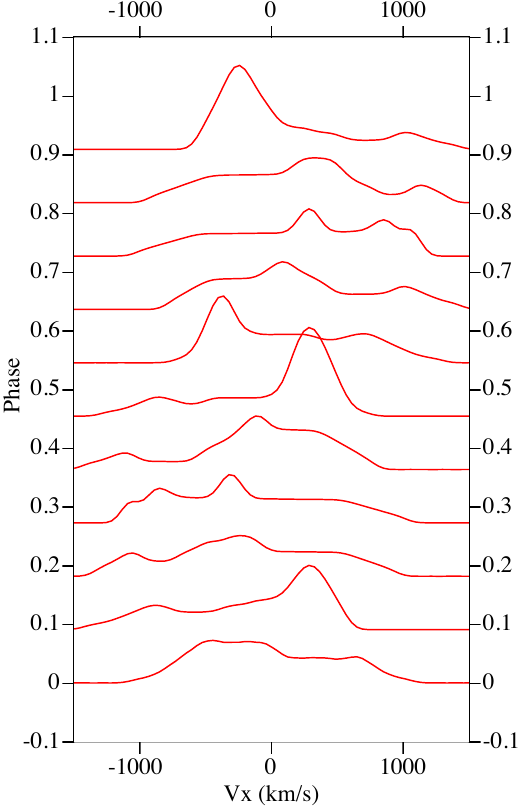}} &
      \raisebox{-\totalheight}{\includegraphics[width=3.5cm]{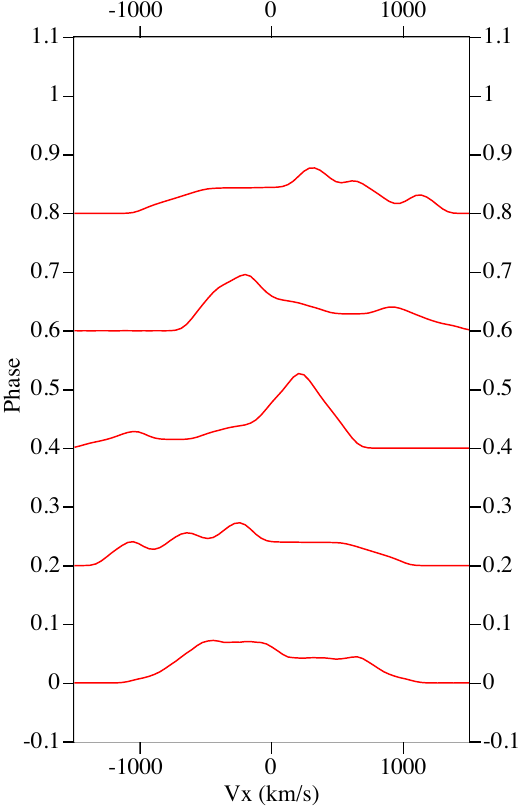}} &
      \raisebox{-\totalheight}{\includegraphics[width=3.5cm]{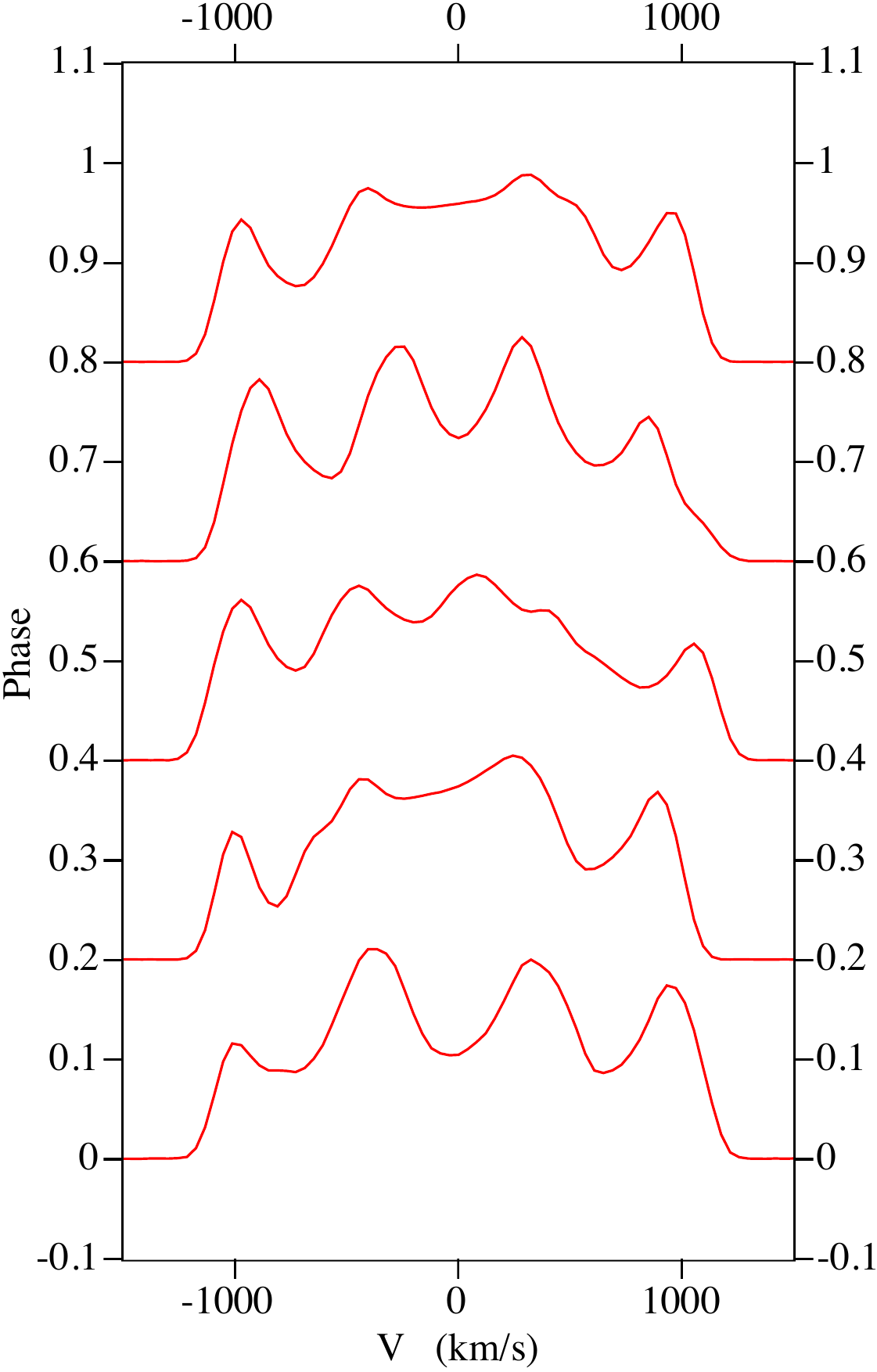}} &
      \raisebox{-\totalheight}{\includegraphics[width=3.5cm]{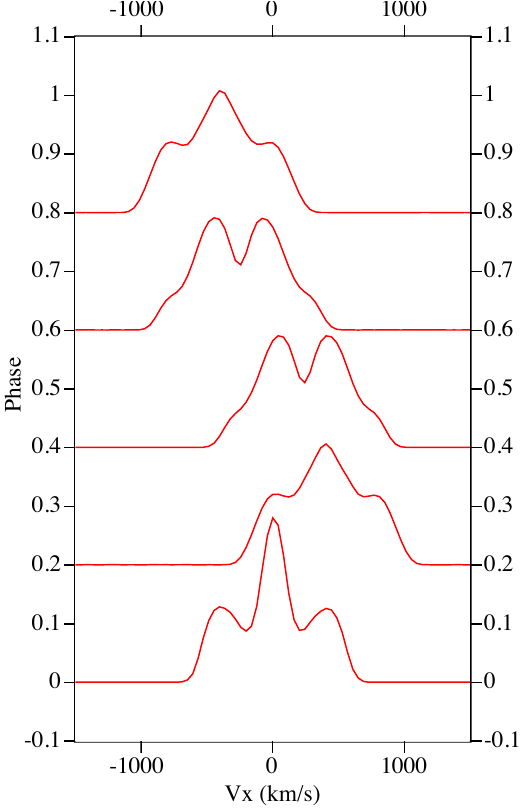}}\\
      \raisebox{-\totalheight}{\includegraphics[width=3.5cm]{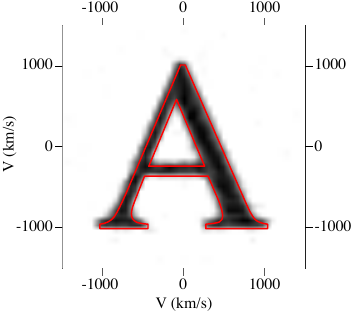}} & 
      \raisebox{-\totalheight}{\includegraphics[width=3.5cm]{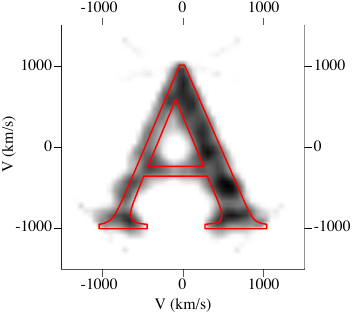}} & 
      \raisebox{-\totalheight}{\includegraphics[width=3.5cm]{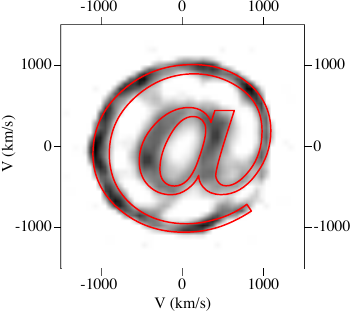}} & 
      \raisebox{-\totalheight}{\includegraphics[width=3.5cm]{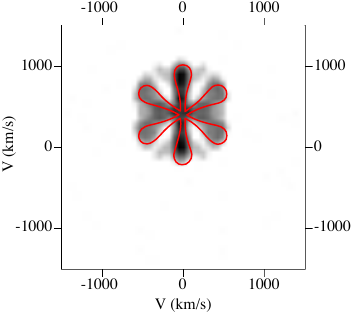}}\\
      (a) & (b) & (c) & (d)\\
    \end{tabular}
  \end{center}
\caption{Reconstruction results for (a) the symbol ``A'' by 11 profiles, (b) the symbol ``A'' by five profiles, (c) the symbol ``@'' by five profiles, and (d) the symbol ``*'' by five profiles, while the remaining parameters are similar to those in Fig.~\ref{letterA1}.}\label{lowres}
\end{figure}

The restoration of the image of the letter ``A'' using 11 and five profiles, as well as the symbols ``@'' and ``*'' using five profiles is shown in Fig.~\ref{lowres}. As can be seen from the obtained results, the algorithm copes very well with recovery based on a small number of profiles. In the case of 11 profiles, only weak artifacts can be seen in the reconstructed image in the area of the ``legs'' of the letter ``A''. For five profiles, artifacts are significantly more numerous (especially for ``*''). However, all the
main image elements are reconstructed with good correspondence to the original image.

\subsection{Recovering Blurred Data}

\begin{figure}[ht]
  \begin{center}
	\begin{tabular}{ccccc}
      \raisebox{-\totalheight}{\includegraphics[width=2.75cm]{figs/letterAP21W300S1000/blured.pdf}} &
      \raisebox{-\totalheight}{\includegraphics[width=2.75cm]{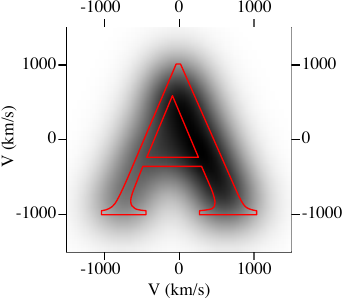}} &
      \raisebox{-\totalheight}{\includegraphics[width=2.75cm]{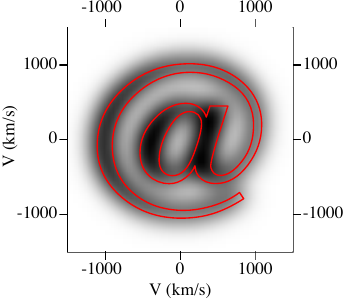}} &
      \raisebox{-\totalheight}{\includegraphics[width=2.75cm]{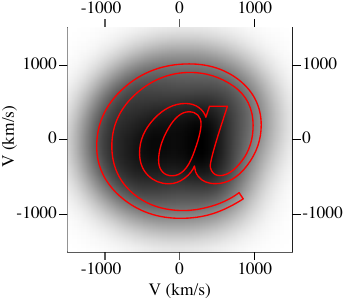}} &
      \raisebox{-\totalheight}{\includegraphics[width=2.75cm]{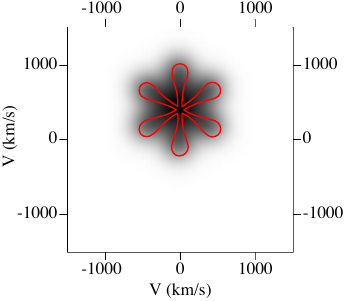}}\\
      \raisebox{-\totalheight}{\includegraphics[width=2.75cm]{figs/letterAP21W300S1000/result.pdf}} &
      \raisebox{-\totalheight}{\includegraphics[width=2.75cm]{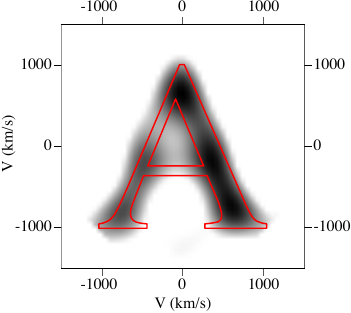}} &
      \raisebox{-\totalheight}{\includegraphics[width=2.75cm]{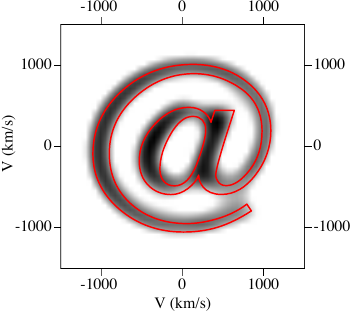}} &
      \raisebox{-\totalheight}{\includegraphics[width=2.75cm]{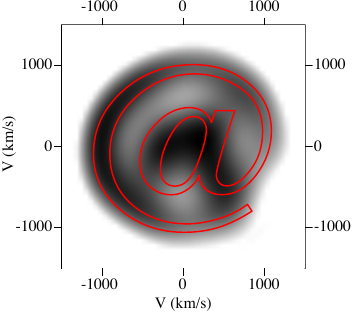}} &
      \raisebox{-\totalheight}{\includegraphics[width=2.75cm]{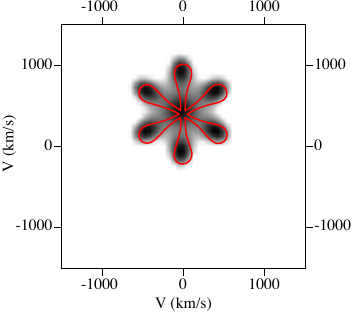}}\\
      \raisebox{-\totalheight}{\includegraphics[width=2.75cm]{figs/letterAP21W300S1000/trail.pdf}} &
      \raisebox{-\totalheight}{\includegraphics[width=2.75cm]{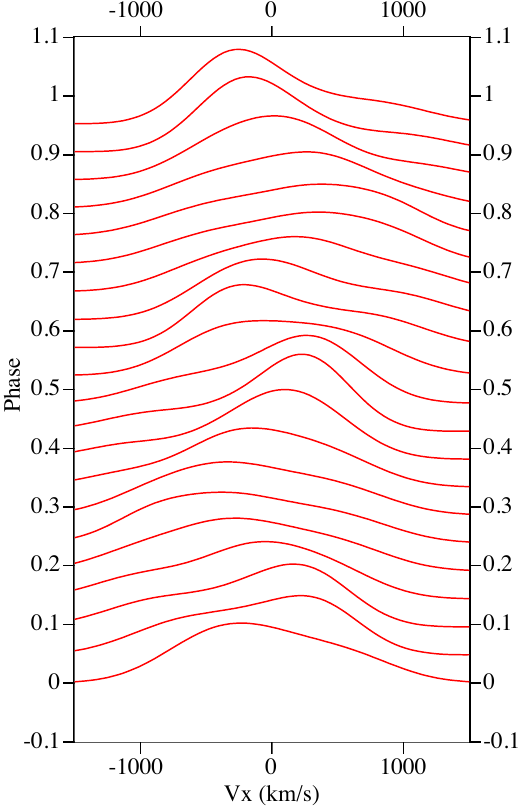}} &
      \raisebox{-\totalheight}{\includegraphics[width=2.75cm]{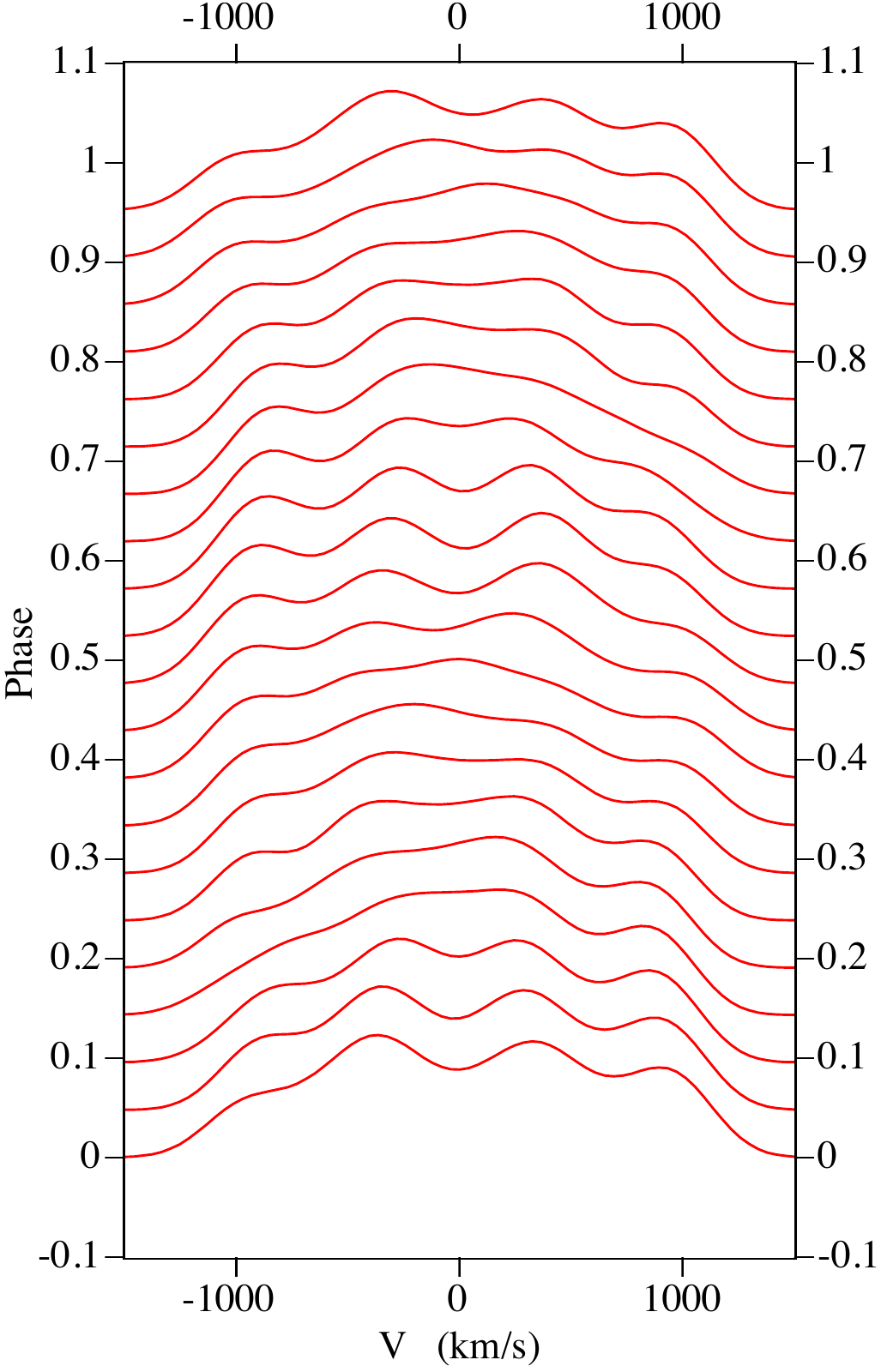}} &
      \raisebox{-\totalheight}{\includegraphics[width=2.75cm]{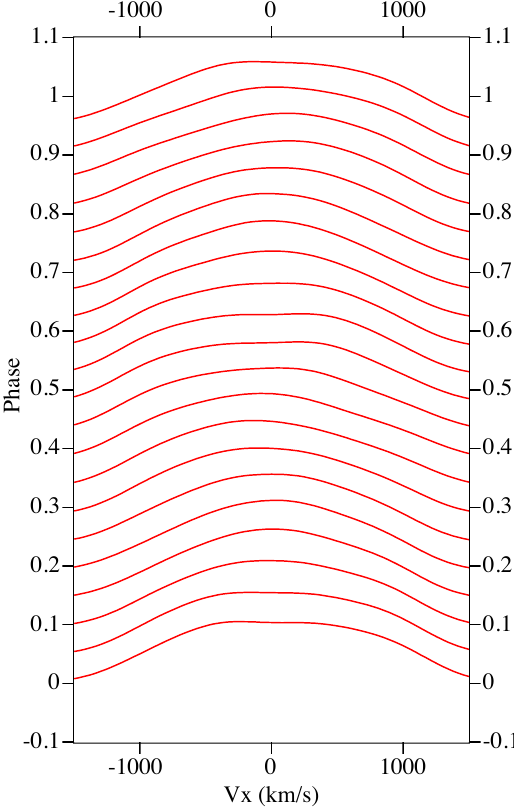}} &
      \raisebox{-\totalheight}{\includegraphics[width=2.75cm]{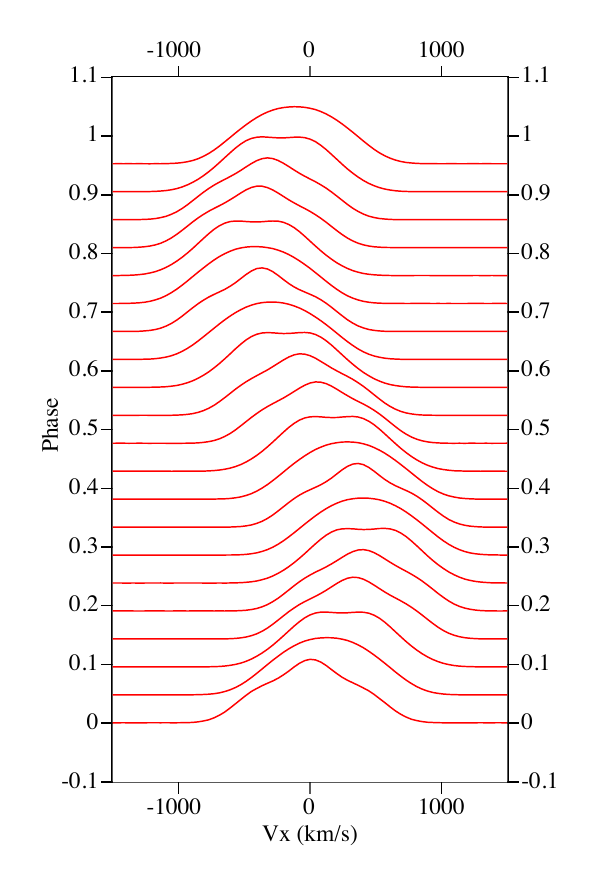}}\\
      (a) & (b) & (c) & (d) & (e)
    \end{tabular}
  \end{center}
\caption{Image reconstruction from blurred profiles: FWHM = 300 km/s (a, c, e, 500 iterations) and 600 km/s (b, d, 1000 iterations). The first row is the initial images (convolved with a Gaussian with the corresponding FWHM), the second row is the reconstructed images, and the third row is the profiles used for reconstruction.}\label{blur}
\end{figure}

The image recovery with greater blur than that in Fig.~\ref{letterA1} is shown in Fig.~\ref{blur} with FWHM = 300 km/s (columns a, c and d, 500 iterations) and FWHM = 600 km/s (columns b and d, 1000 iterations). The algorithm well coped with removing blur at FWHM = 300, the general contours of the image were restored with good accuracy, only in sharp corners and in areas, where the gap between elements is very small, some smoothness appeared. Even the bridge of the letter ``A'' and the outer circle of the ``@'' symbol were well restored, although some non-uniformity was observed in them. When restoring the ``*'' symbol on the oroginal blurred image, individual rays are almost invisible, however, the algorithm was able to restore the size and location of their ends.

In the reconstructed image of the letter ``A'', at FWHM = 600, the wide diagonal line has a width that is almost identical to the original, while the narrow diagonal line is noticeably blurred (approximately twice the width) and non-uniform. The bridge has disappeared, and instead, there are two thickenings on the diagonal lines, forming a ``gap'' area, which is also very blurred. The serifs at the bottom of the letter also turned into two thickenings. However, the image matches the initial very well, considering the extreme degree of blurring of the initial data. For the ``@'' symbol, the algorithm was able to reconstruct the gap between the letter and the line surrounding it, albeit very blurry, but the gap within the letter ``a'' itself was not resolved.

The calculation was artificially stopped after 500 (FWHM = 300) and 1000 (FWHM = 600) iterations. However, by increasing the number of iterations, it would be possible to achieve higher quality images. The algorithm’s computation speed has an inverse square dependence on the FWHM value, so that at FWHM = 300, for example, each iteration took nine times longer than that at FWHM = 100. However, the task remains adequately feasible for a personal computer, and 500 iterations took less than half an hour on an eight-year-old laptop.

\subsection{Reconstruction From Noisy Data}

\begin{figure}[p]
  \begin{center}
	\begin{tabular}{c|ccccc|}
      \multicolumn{1}{c}{\multirow{10}{1.5cm}{\raisebox{-8.5cm}{\rotatebox{90}{$m$}}}}&\multicolumn{5}{c}{}\\
      \multicolumn{1}{c}{}&\multicolumn{5}{c}{SNR}\\
      \multicolumn{1}{c}{}&
      {\small 1000} & {\small 10} & {\small 5} & {\small 2} & \multicolumn{1}{c}{\small 1}\\
      \cline{2-6}&&&&&\\
      \raisebox{1cm}{1.0}&
      \includegraphics[width=2.5cm]{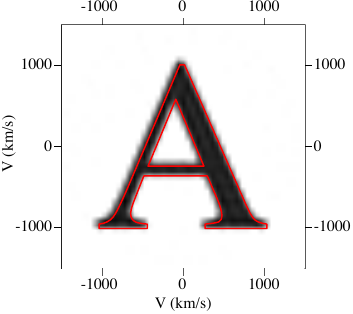} &
      \includegraphics[width=2.5cm]{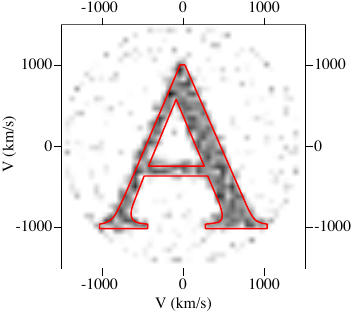} &
      \includegraphics[width=2.5cm]{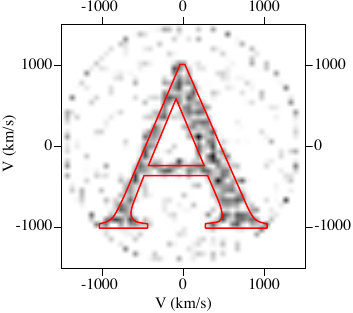} &
      \includegraphics[width=2.5cm]{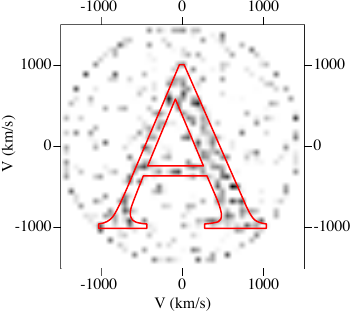} &
      \includegraphics[width=2.5cm]{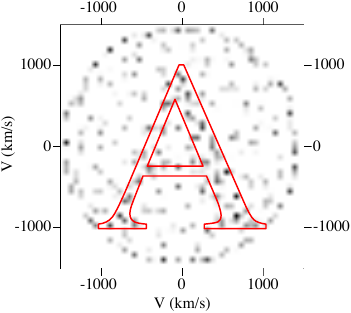}\\
      \raisebox{1cm}{0.99} &
      \includegraphics[width=2.5cm]{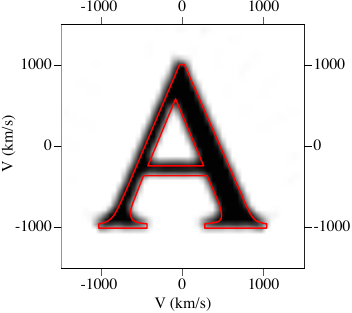} &
      \includegraphics[width=2.5cm]{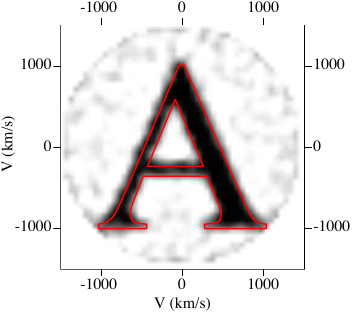} &
      \includegraphics[width=2.5cm]{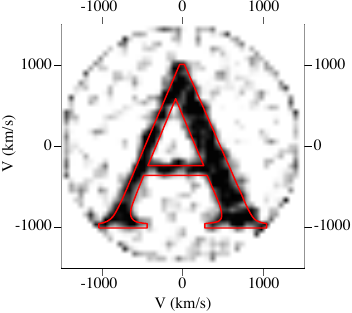} &
      \includegraphics[width=2.5cm]{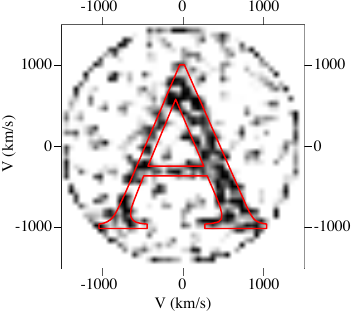} &
      \includegraphics[width=2.5cm]{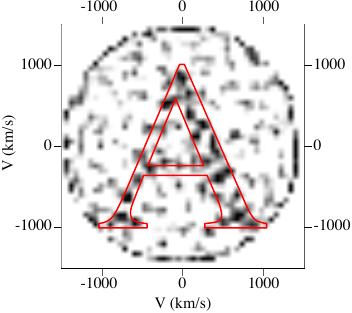}\\
      \raisebox{1cm}{0.95} &
      \includegraphics[width=2.5cm]{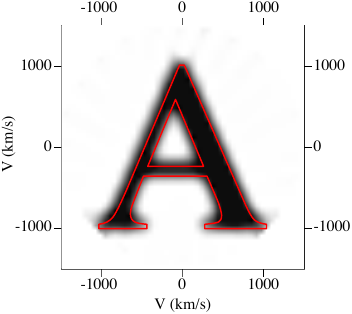} &
      \includegraphics[width=2.5cm]{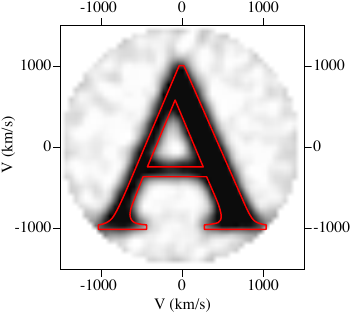} &
      \includegraphics[width=2.5cm]{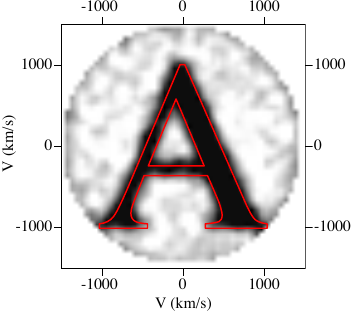} &
      \includegraphics[width=2.5cm]{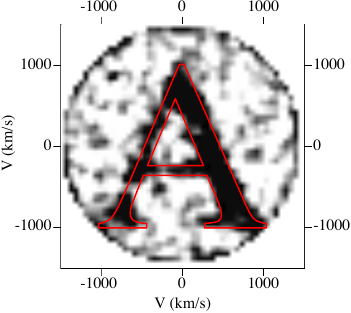} &
      \includegraphics[width=2.5cm]{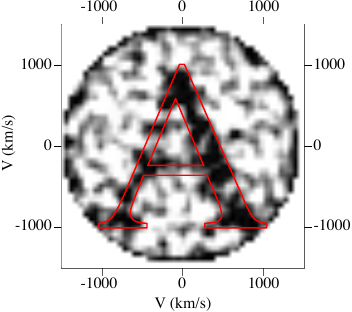}\\
      \raisebox{1cm}{0.9} &
      \includegraphics[width=2.5cm]{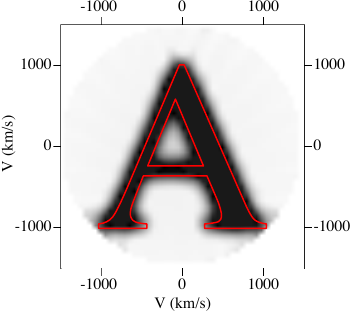} &
      \includegraphics[width=2.5cm]{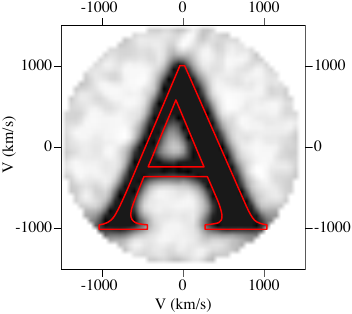} &
      \includegraphics[width=2.5cm]{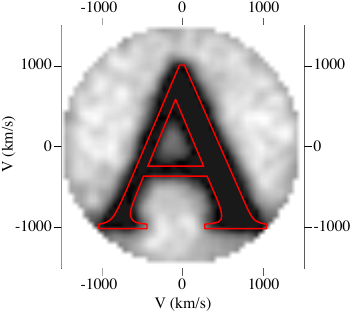} &
      \includegraphics[width=2.5cm]{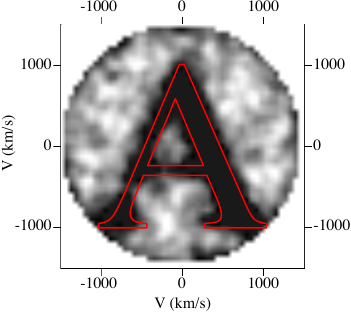} &
      \includegraphics[width=2.5cm]{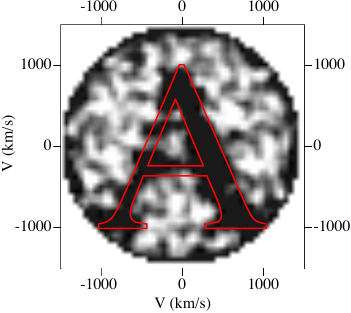}\\
      \raisebox{1cm}{0.8} &
      \includegraphics[width=2.5cm]{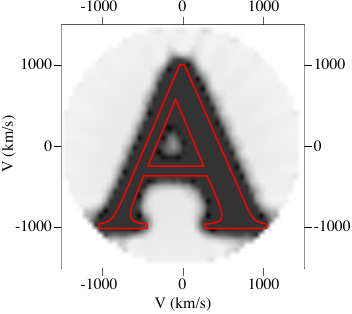} &
      \includegraphics[width=2.5cm]{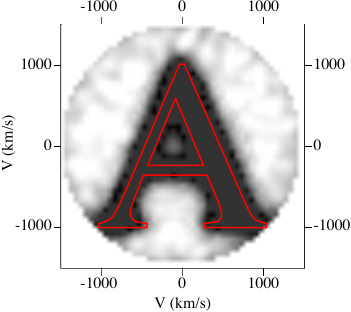} &
      \includegraphics[width=2.5cm]{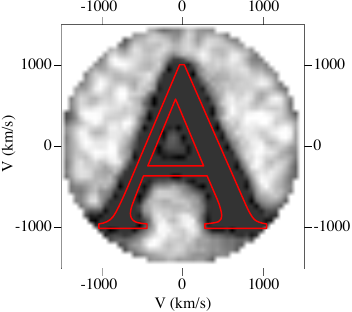} &
      \includegraphics[width=2.5cm]{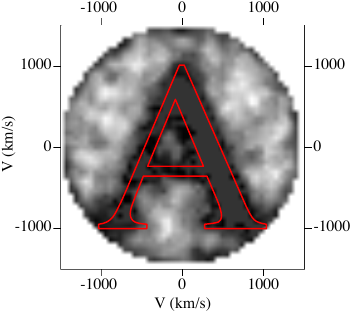} &
      \includegraphics[width=2.5cm]{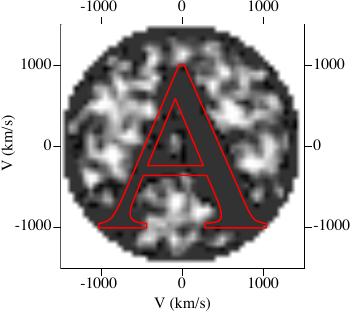}\\
      \cline{2-6}&&&&&\\
      \raisebox{2cm}{\rotatebox[origin=c]{90}{profiles}}&
      \includegraphics[width=2.5cm]{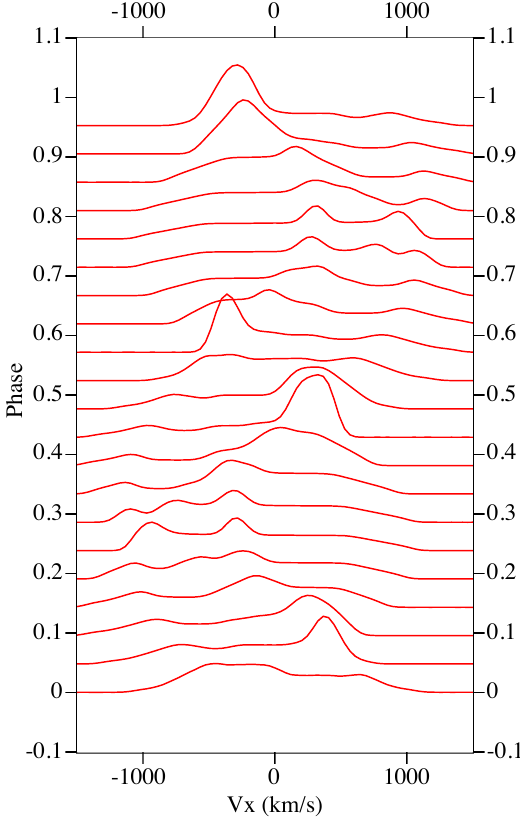} &
      \includegraphics[width=2.5cm]{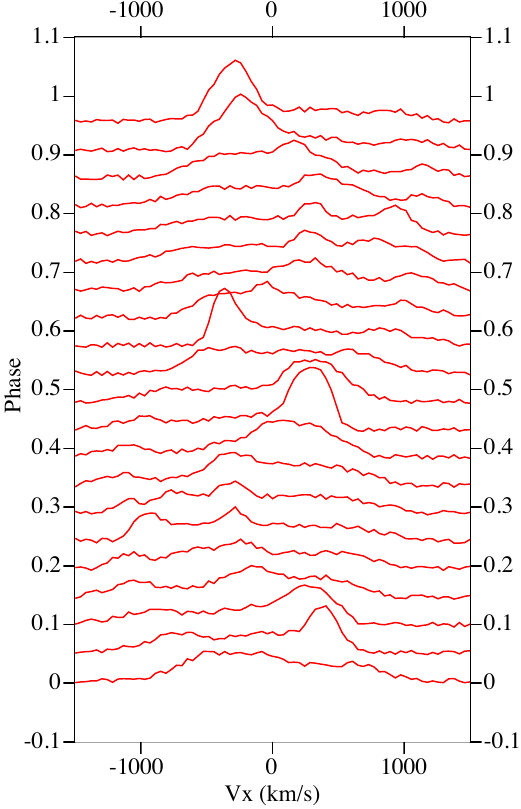} &
      \includegraphics[width=2.5cm]{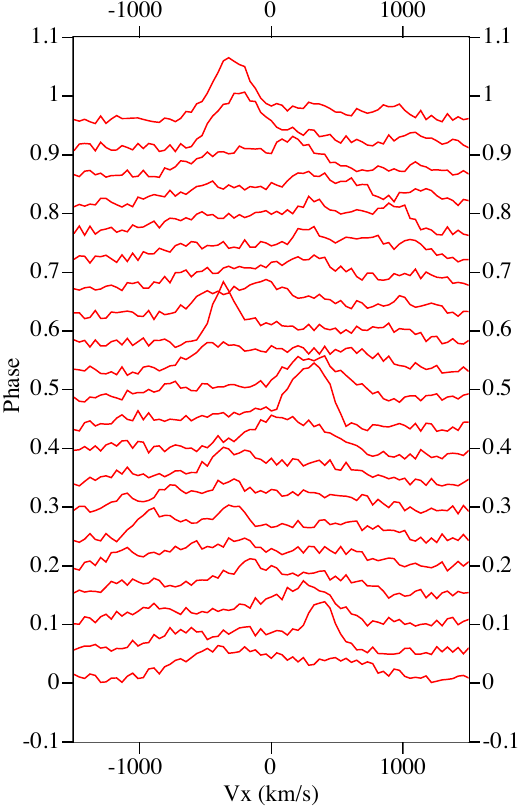} &
      \includegraphics[width=2.5cm]{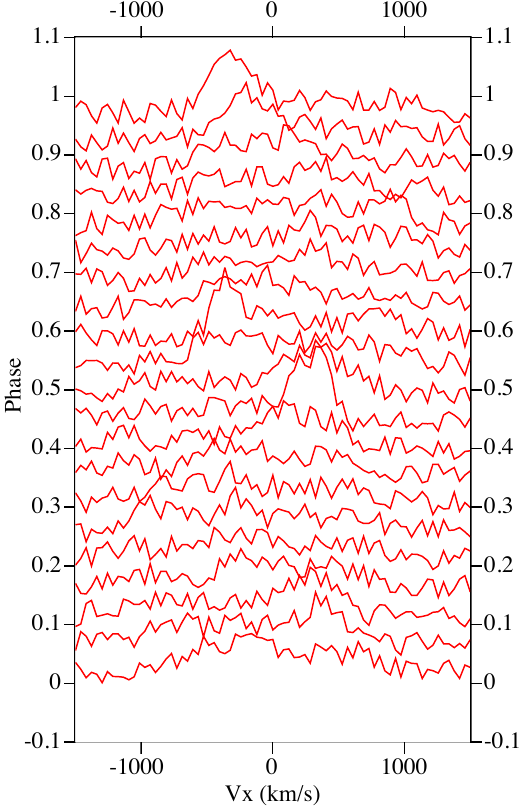} &
      \includegraphics[width=2.5cm]{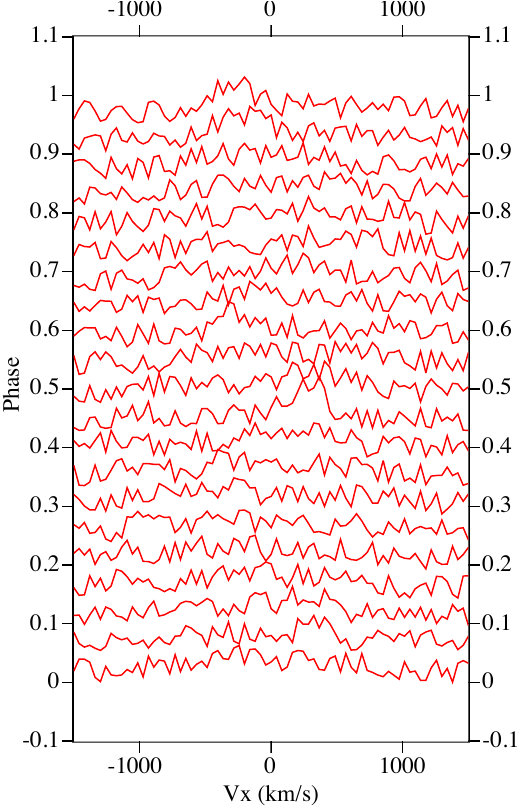}\\
      \cline{2-6}
    \end{tabular}
  \end{center}
\caption{Reconstruction of noisy images at FWHM = 100, 21 profiles, SNR = 1000, 10, 5, 2 and 1, and with different cutoff threshold values.}\label{ASNR}
\end{figure}

The ART method operates well with noise-free data, but in the presence of significant noise, simple minimization results in the image being broken up into many individual points separated by empty space. This happens because noisy profiles have a sawtooth shape, with many local maxima that the algorithm tries to reproduce in the image. If the noise power exceeds a certain threshold (according to test results, this occurs at a signal-to-noise ratio or SNR of ~3–5 and below), where SNR is defined as the ratio of the root-meansquare signal values minus the DC component to the root-mean-square noise level. If the SNR is sufficiently small, local noise spikes contribute more to the computation of $\chi^2$, than the useful signal, so that the initial image appears less bright than the peaks obtained by approximating the noise. In the development of Tomo-V, several different approaches were tested to reduce the impact of noise, and ultimately, a method that essentially was reduced to finding a local minimum characterized by a smaller number of outliers was implemented. When using this method, a certain threshold, $m<1$, is set, and, at each step of minimization, the brightness of all points in the image whose brightness is less $m\cdot I_{max}$, where $I_{max}$ is the maximal brightness in the image that is multiplied by $m$. This ``disrupts'' the minimization process, allowing it to jump out of the ``ravine'' leading to the minimum that produces a ``crumbling'' image, and find a local minimum that is less optimal, but also characterized to a lesser extent by the loss of the main signal.

The reconstruction results for noisy source data at various signal-to-noise ratios (SNRs) and different values are shown in Fig.~\ref{ASNR}. For all images, FWHM = 100 km/s was taken. Signal noise is produced by adding white noise, a random variable whose root mean square value is related to the root mean square value of the signal as 1/SNR. The top row of images in Fig.~\ref{ASNR} corresponds to the threshold value of $m=1$, that is, it is not cut. For these cases, the number of steps was limited to 500. For relatively weak noise (SNR = 10), an image of sufficiently good quality, from which the parameters of all elements can be determined with good accuracy, is obtained. The image differs from the ``ideal'' case (SNR = 1000) by the presence of noise in both light and dark areas. At SNR = 5, the picture is about the same, except for more noise. In the case of SNR = 2, the image contours are practically no longer distinguished, as at SNR = 1.

Enabling a small cut (second line of Fig.~\ref{ASNR}, $m=0.99$) has virtually no effect on the recovery of ideal data (first column), but provides significant improvement in images in the presence of noise. For SNR = 10, a very clear image of the initial letter without artifacts in the black areas is obtained, but there is some number of weak artifacts in the background. For SNR = 5, a clear image was also obtained, with slightly more pronounced background noise. At SNR = 2 the image has artifacts both in the background and in black areas, but the letter contours are defined very well. For SNR = 1, there are also improvements, but the letter outlines are still difficult to discern. In these cases, the local minimum is reached in a smaller number of steps, $\sim 40$ for SNR = 1000 and $\sim 10$, and $\sim 100$ for the rest.

Stronger cutting (third line of Fig.~\ref{ASNR}, $m=0.95$) already has a small impact on the reconstruction of ideal data in the first column; the image turns out a little blurry. The same can be said about the image for SNR = 10 and 5. In case of SNR = 2, a significantly sharper image was obtained compared to $m=0.99$, with almost no artifacts in black areas, but background noise became more noticeable. For data with SNR = 1 (right column), a significant improvement was achieved: the letter contours are now clearly visible, but the background noise has also become stronger. In this case, the minimization reaches a local minimum in 20-40 steps.

Cutting with threshold of $m=0.9$ (fourth row of Fig.~\ref{ASNR}) already leads to a noticeable deterioration for ideal data (left column), the image is more blurred and the background is no longer white; for SNR = 10, 5, and 2, the picture is similar. At SNR = 1, the image of the letter itself is obtained almost without artifacts, but the background noise is very strong, the spots are comparable in intensity to the main image. The local minimum is reached in 15–20 steps. Further reduction of the threshold to (fifth row of Fig.~\ref{ASNR}) no longer provides any improvement even for SNR = 1, leading only to stronger blurring and an increase in background noise.

\subsection{Reducing Noise in Initial Profiles}

Typically, to increase the signal-to-noise ratio, the signal accumulation time is increased, essentially averaging the noisy signal over a longer period of time. This method has limited applicability when obtaining data for Doppler tomography of short-period objects, since the phase of the system must not change significantly during the accumulation time. As a result, the data represent a compromise between the desire to obtain less noisy data and the need to have sufficiently good coverage across the phases of the system. However, to some extent, averaging can be carried out using existing profiles, since a considerable part of the information in them is duplicated. For example, if we have two profiles taken at exactly opposite phases, their shape must be identical, but mirrored relative to zero velocity. So that, by mirroring one of the profiles, we can average them, obtaining a single profile with half the signal-to-noise ratio.

In reality, it is very rare case when two profiles are taken at exactly opposite phases. Obtaining such profiles is not practical, since it does not improve the coverage by phases; they will be equivalent to one profile. However, if there are many profiles, usually for each of them, a pair located on either side near the opposite phase can be found, and from these, the desired profile exactly corresponding to the opposite phase can be obtained by interpolation. In interpolation, the Central Projection Theorem, which states that the Fourier transform of any profile corresponds to a slice of the two-dimensional Fourier transform of the reconstructed image, can be used. Thus, it is possible to interpolate between the coefficients of the Fourier images for two adjacent profiles in order to obtain the corresponding coefficients of the Fourier transform of the desired profile at the desired phase, and then, using the inverse transformation, to obtain the profile itself.

\begin{figure}[t]
\centering
\begin{tabular}{cccc}
  \raisebox{1.2cm}{\rotatebox{90}{m=1.0}} &
  \includegraphics[width=4.2cm]{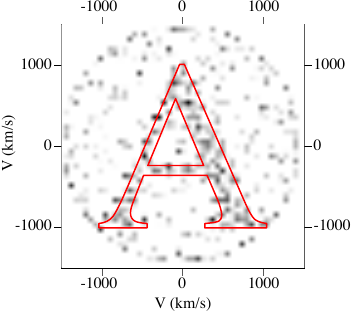} &
  \includegraphics[width=4.2cm]{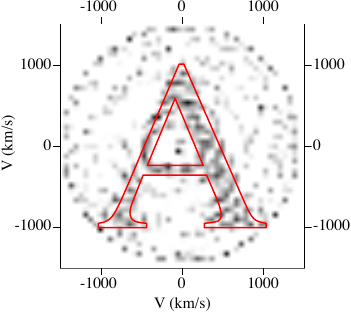} &
  \includegraphics[width=4.2cm]{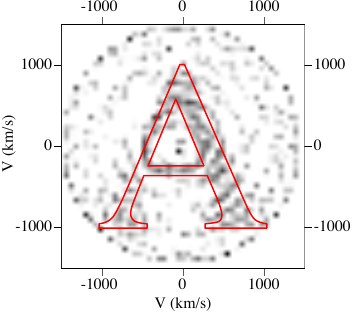}\\
  \raisebox{1.2cm}{\rotatebox{90}{m=0.97}} &
  \includegraphics[width=4.2cm]{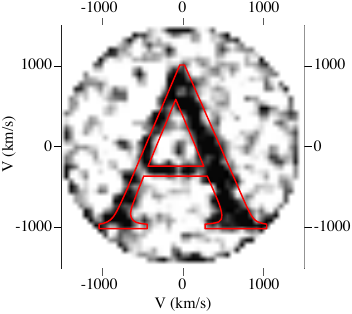} &
  \includegraphics[width=4.2cm]{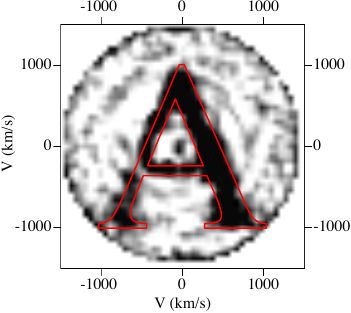} &
  \includegraphics[width=4.2cm]{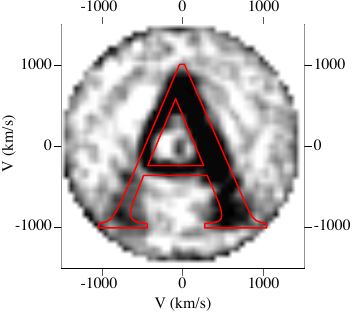}\\
  \raisebox{2.5cm}{\rotatebox{90}{profiles}} &
  \includegraphics[width=4.2cm]{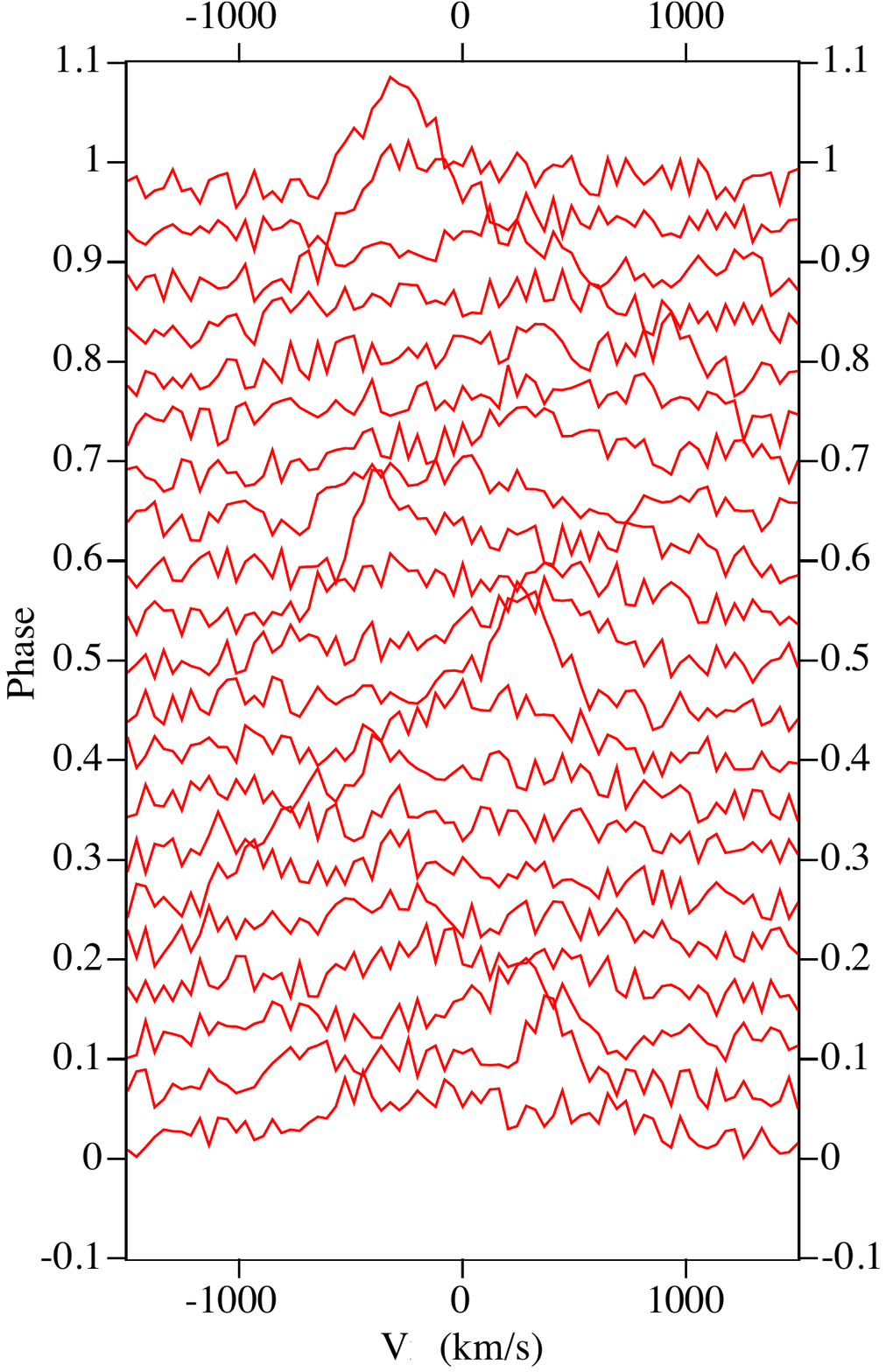} &
  \includegraphics[width=4.2cm]{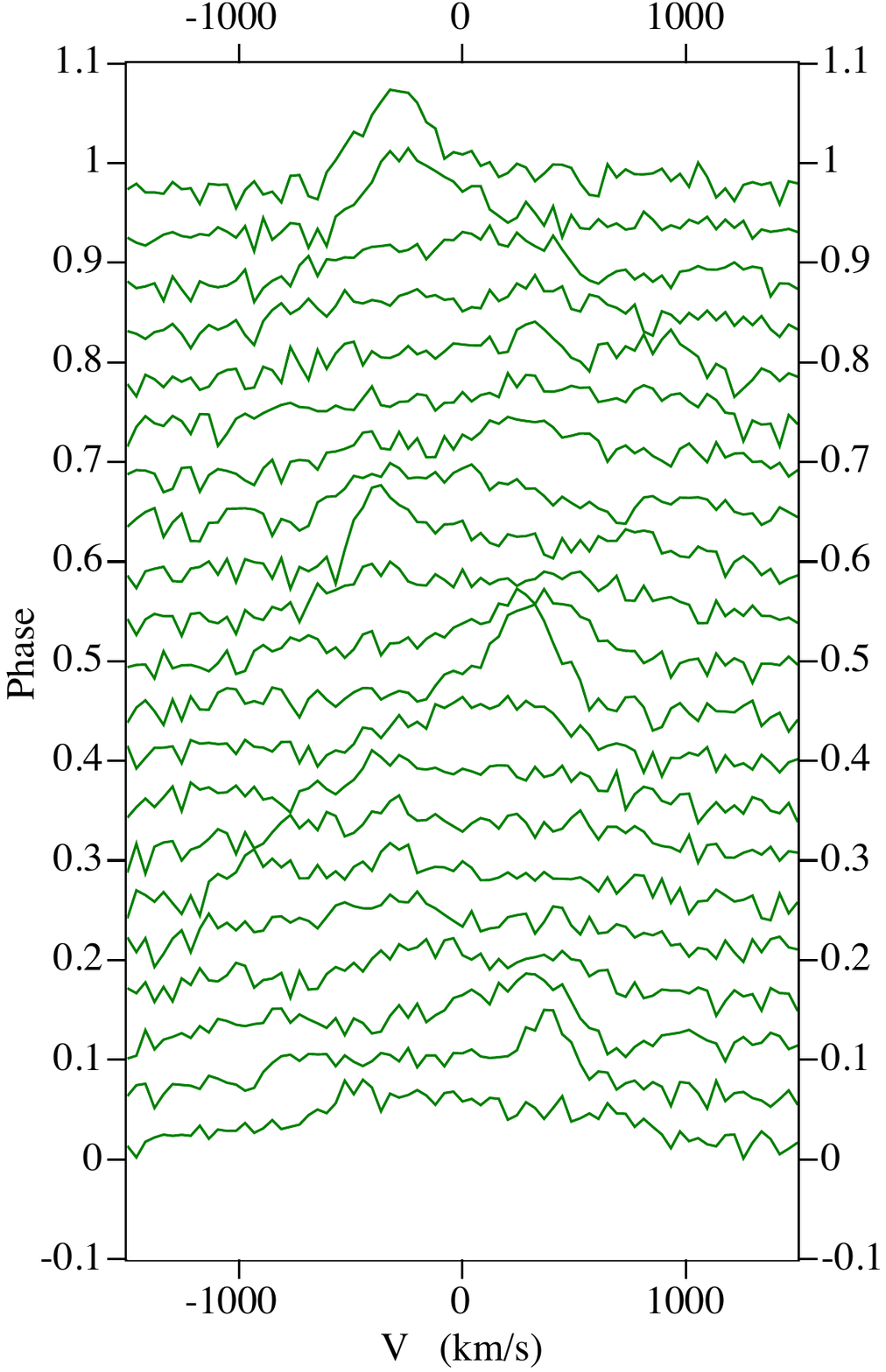} &
  \includegraphics[width=4.2cm]{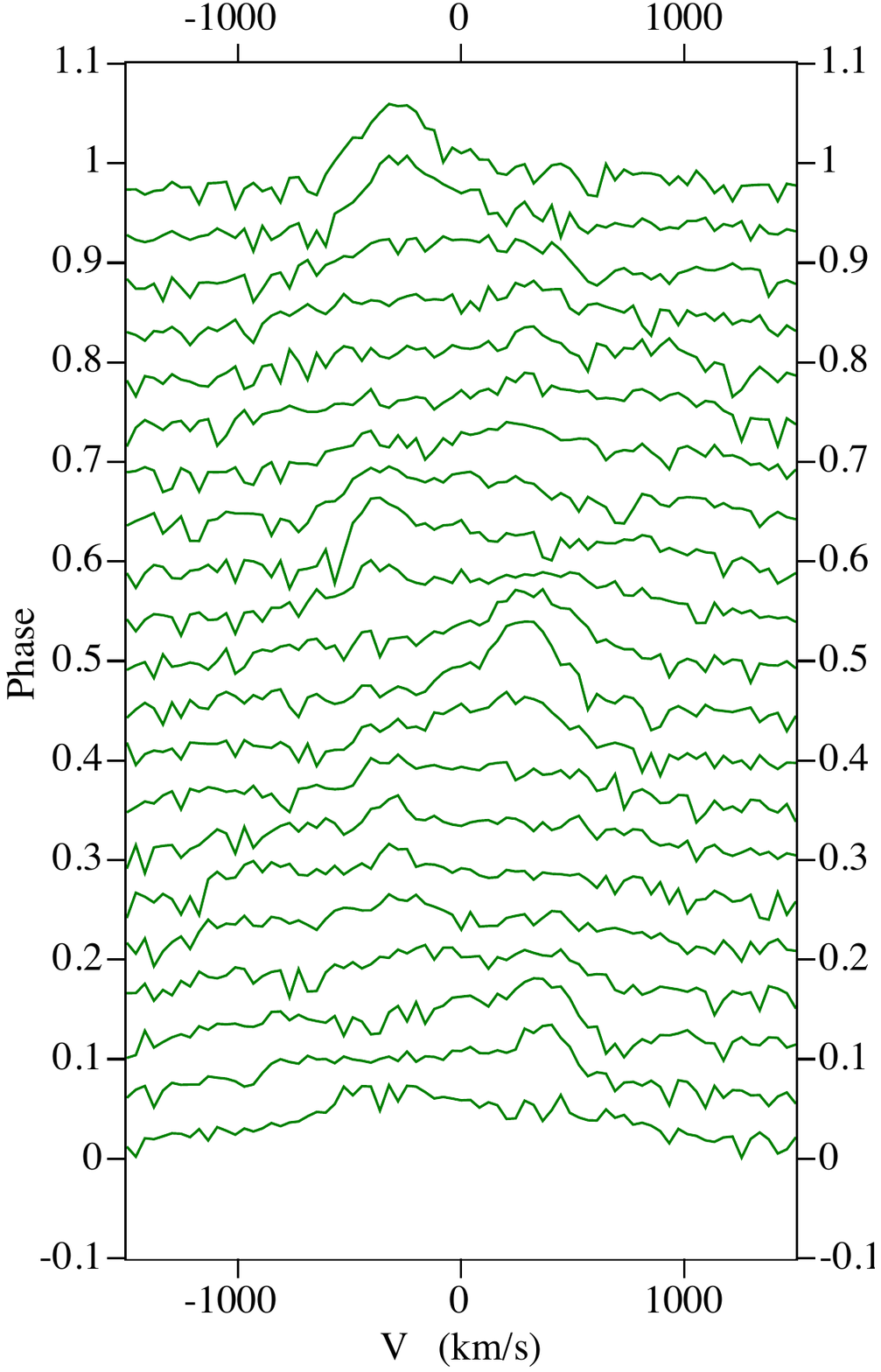}\\
  &(a) & (b) & (c)
\end{tabular}
\caption{Image reconstruction from profiles (a) without averaging, (b) averaged with opposite-phase profiles, and (c) averaged with opposite and nearby profiles. SNR = 2, 21 profiles, and FWHM = 100.}
\label{denoise1}
\end{figure}

In Tomo-V, such averaging can be performed by selecting the appropriate option in the ``Denoise'' selector, below the trailed spectra image. When selecting ``using opposite profiles'', averaging for each profile will be performed using the interpolated profile at the opposite phase. When selecting ``using opposite and closest profiles'' averaging will use the profile interpolated over the two adjacent ones. Averaging for each specific profile will be performed if the phase difference between the two profiles taken for interpolation is less than 0.25, which makes it possible to avoid too strong artifacts due to incomplete phase coverage.

The results of reconstruction using synthetic profiles with an SNR ratio of 2, without averaging (column a), with averaging of each profile with an interpolated profile opposite in phase (column b), and also with opposite and adjacent profiles (column c) are shown in Fig.~\ref{denoise1}. The first line shows the results of recovery without using peak clipping ($m=1.0$), in the second, with cutting at $m=0.97$, and the third column shows the profiles themselves. As can be seen from the figure, averaging improves image quality to some extent, but also leads to ``smearing'' of background artifacts.

\begin{figure}[t]
\centering
\begin{tabular}{cc}
  \includegraphics[width=5cm]{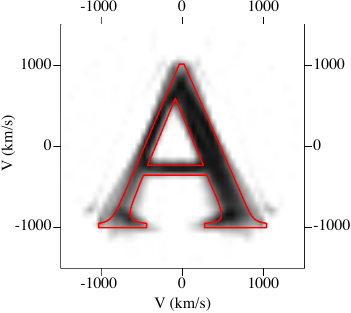} &
  \includegraphics[width=5cm]{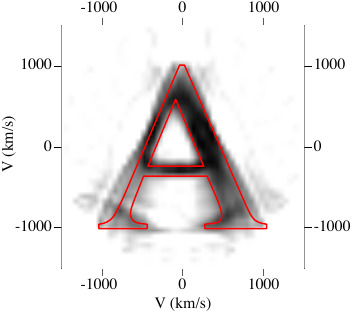}\\
  \includegraphics[width=5cm]{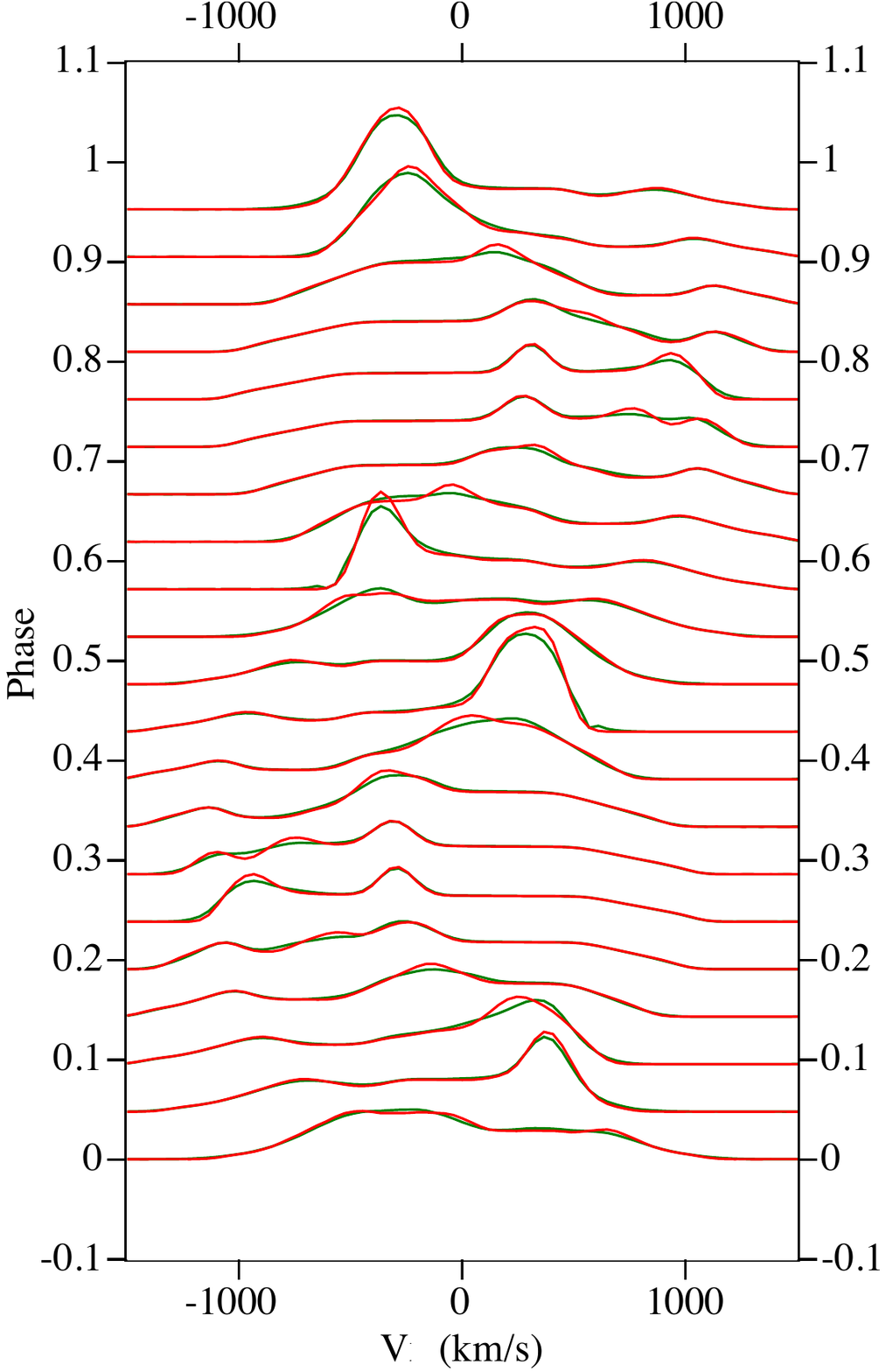} &
  \includegraphics[width=5cm]{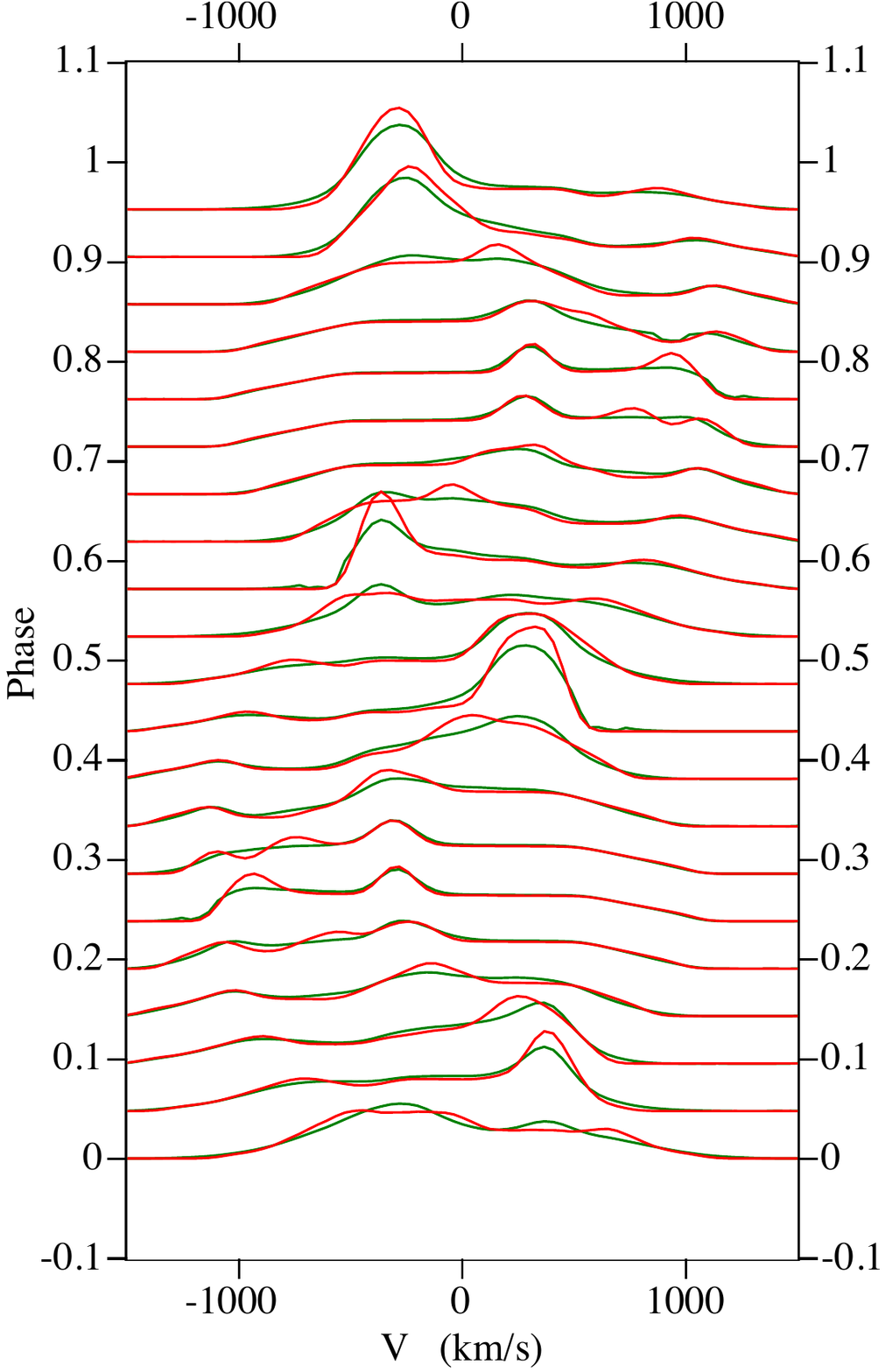}\\
  (a) & (b)
\end{tabular}    
\caption{Artifacts appearing during reconstruction with averaging when (a) using opposite-phase profiles, (b) as well as opposite and nearby profiles. The lower panels show the initial (red) and averaged (green) profiles.}
\label{denoise2}
\end{figure}

The reconstruction results for noise-free data when averaging with opposite-phase profiles (column a), as well as with opposite and adjacent profiles (column b), are shown in Fig.~\ref{denoise2}. Also, the second row in Fig.~\ref{denoise2} shows the initial profiles (red lines) superimposed on the averaged profiles (green lines). As can be seen from the figures, averaging results in a small change in the shape of the profiles, mainly noticeable in the peak region. In this case, artifacts appear on the images as ``shadows'' noticeable near the protrusions in the image. When using averaging across both opposite and adjacent profiles, the image loses some fine details.

\section{Testing on Real Data}

To test on real data, profiles in the $\mathrm{H}_\beta$ line obtained in 2006 for the star SS Cyg, which is in a state of outburst~\cite{2008ARep...52..835K} were taken. SS Cyg is a semi-detached binary star consisting of a red dwarf with a mass of$\sim 0.56\mathrm{M}_\odot$\footnote{From here on, the parameters of SS Cyg used in~\cite{2008ARep...52..835K} are adopted} that fills its Roche lobe and a white dwarf with a mass $\sim 0.97\mathrm{M}_\odot$ of that accretes matter from the secondary component. The orbital inclination is $i=40^\circ$ relative to the line of sight. In~\cite{2008ARep...52..835K}, a tomogram in the $\mathrm{H}_\beta$ line was not constructed, since it was hampered by the presence of strong absorption in the envelope during the outburst, but with the ART method implemented in Tomo-V, such reconstruction became possible.

\begin{figure}[t]
\centering
\begin{tabular}{cc}
  \multirow[t]{2}{*}{\raisebox{-5.5cm}{\includegraphics[height=11cm]{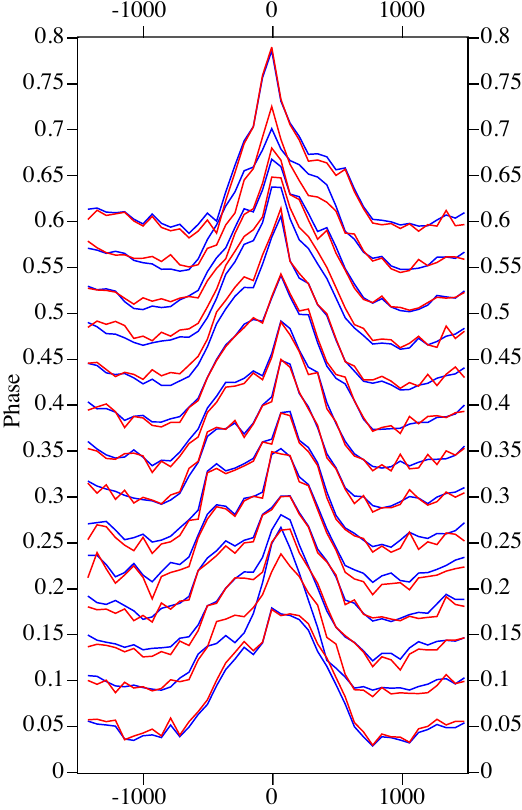}}} &
  \includegraphics[height=5cm]{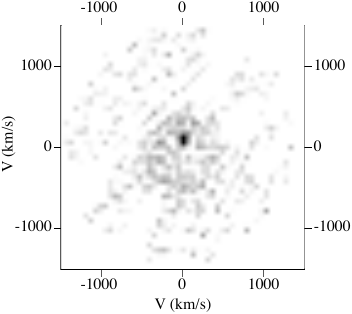}\\
  & \includegraphics[height=5cm]{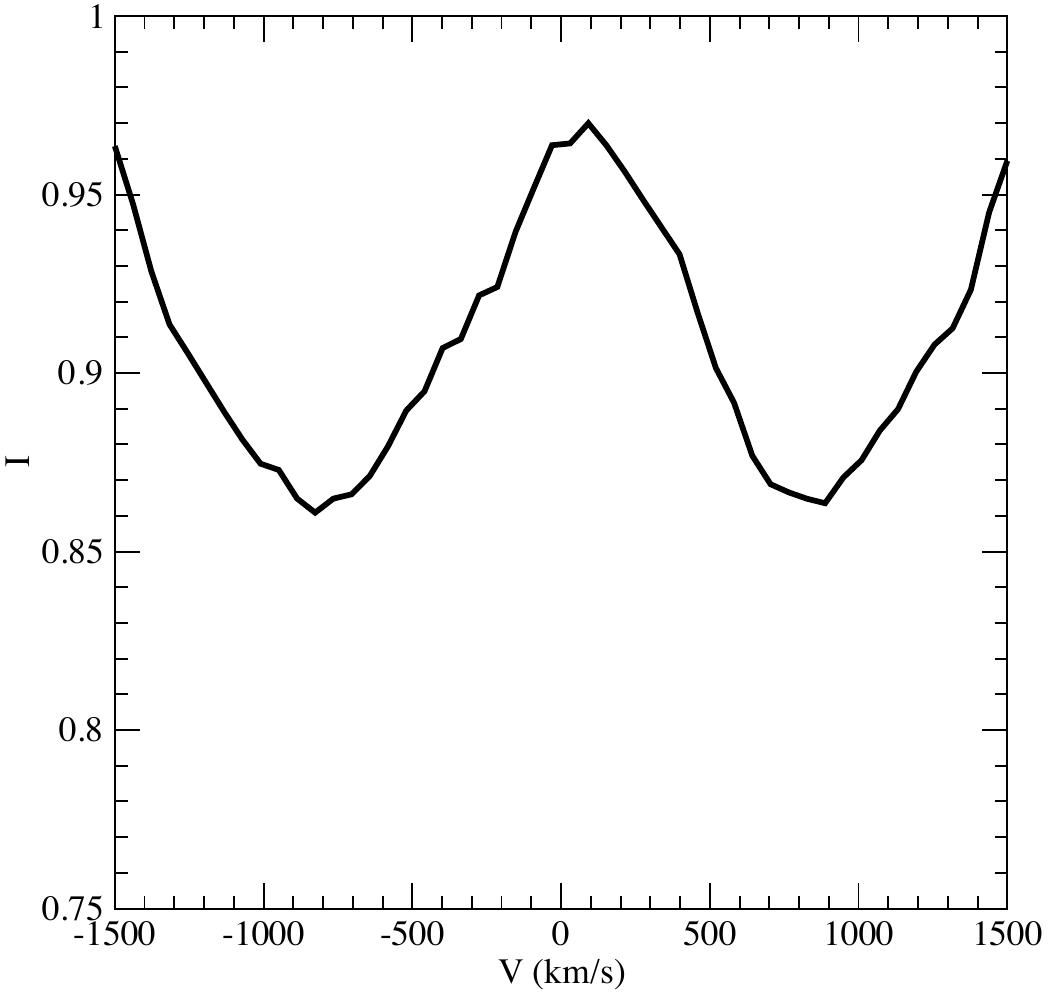}\\
\end{tabular}
\caption{Profiles in the $\mathrm{H}_\beta$ line from~\cite{2008ARep...52..835K} (left column), reconstructed tomogram (right column at the top), and absorption profile (right column at the bottom). The profiles obtained from observations are shown in red, and the synthetic profiles obtained from the reconstructed tomogram are shown in blue. Minimization is stopped after 500 steps.}
\label{SSCygHb1}
\end{figure}

\begin{figure}[t]
\centering
\begin{tabular}{cc}
  \multirow[t]{2}{*}{\raisebox{-5.5cm}{\includegraphics[height=11cm]{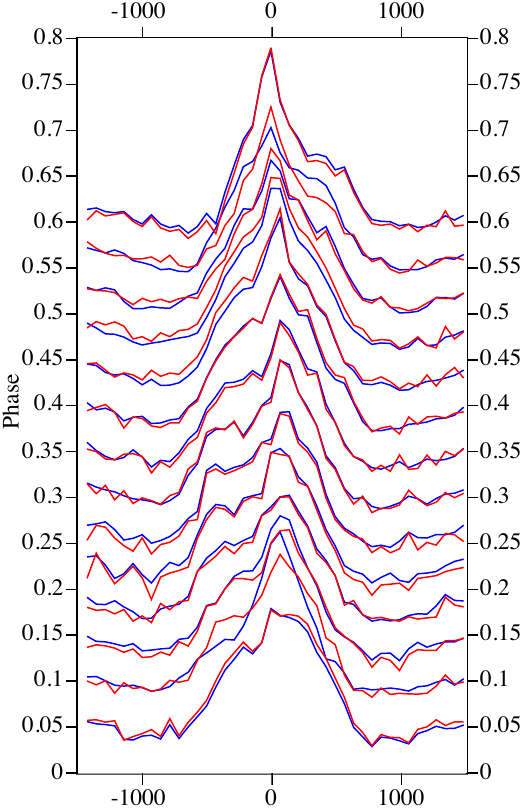}}} &
  \includegraphics[height=5cm]{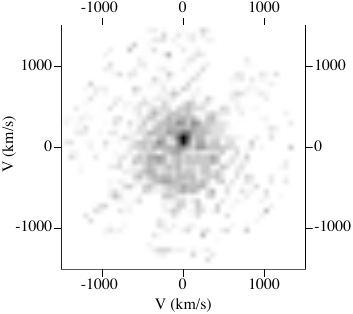}\\
  & \includegraphics[height=5cm]{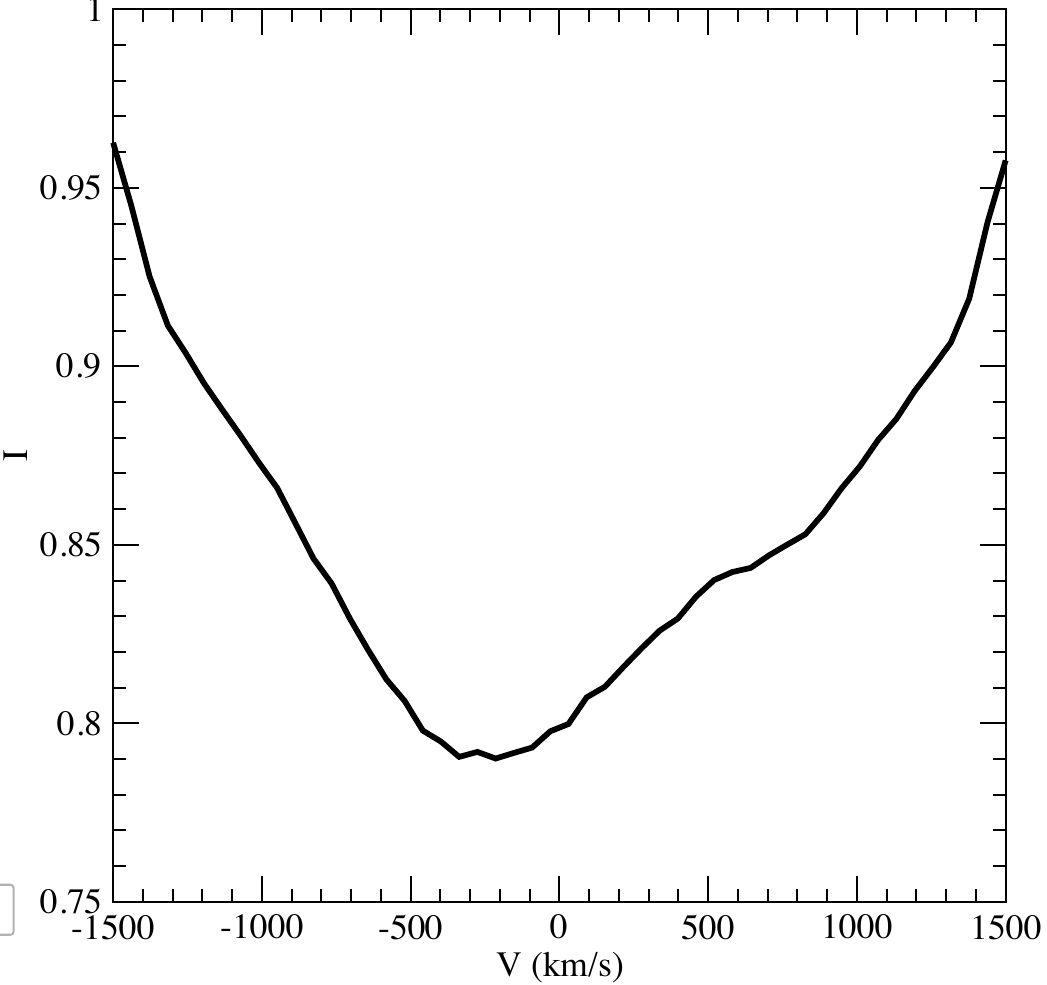}\\
\end{tabular}
\caption{Similar to Fig.~\ref{SSCygHb1}, but the absorption profile is obtained through an intermediate one, in the form of a Gaussian profile.}
\label{SSCygHb2}
\end{figure}

In total, 14 profiles in the $\mathrm{H}_\beta$ line that were obtained on December 13, 2006 were presented in~\cite{2008ARep...52..835K}, the profiles of these lines are shown in Fig.~\ref{SSCygHb1} in red. During the reconstruction, the FWHM (thermal line broadening) parameter was taken to be equal to 100 km/s, since this width was determined for the
main peak in~\cite{2008ARep...52..835K}. Since there is absorption in the line, the ``Fit absorption'' parameter was enabled during reconstruction. As can be seen from the figure, the absorption profile turned out to have two minima, approximately at $\pm 800\,\text{km/s}$. This profile shape may be an artifact, since in the initial data, the emission maximum approximately coincides with the absorption maximum, so that in minimization, two options that fit the input data equally well are possible: with one minimum on the absorption curve, but with a brighter signal on the tomogram in the region of the maximum or with two minima and a less bright signal.

To obtain an absorption curve with one minimum, the reconstruction should be carried out in two stages. At the first stage, the absorption profile should be searched for as a Gaussian profile (for this, the ``Gaussian absorption'' parameter must be enabled), after a certain number of iterations, when the absorption profile stabilizes, the ``Gaussian absorption'' parameter can be disabled and the reconstruction continued until a stable image is obtained. The results of such a two-stage recovery are shown in Fig.~\ref{SSCygHb2}. As can be seen from the coincidence of the observed and synthetic profiles in the left columns of Figs.~\ref{SSCygHb1} and \ref{SSCygHb2}, both options reproduce the observations equally well, so the selection between them must be made based on some physical assumptions.

\begin{figure}[t]
\centering
\begin{tabular}{cccccc}
{\small W} & {\small 20 km/s} & {\small 30 km/s} & {\small 50 km/s} & {\small 70 km/s} & {\small 100 km/s}\\[2mm]
\raisebox{1.2cm}{(a)} &
\includegraphics[width=2.5cm]{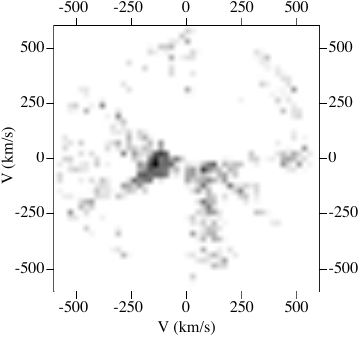} &
\includegraphics[width=2.5cm]{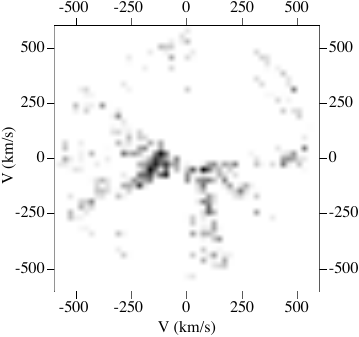} &
\includegraphics[width=2.5cm]{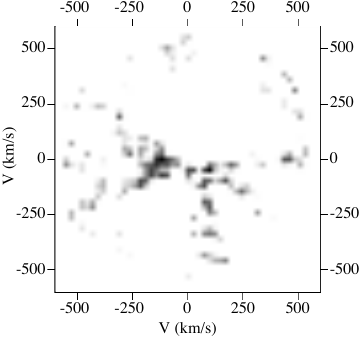} &
\includegraphics[width=2.5cm]{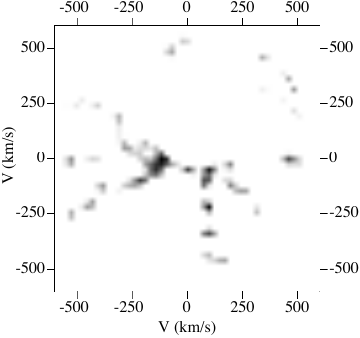} &
\includegraphics[width=2.5cm]{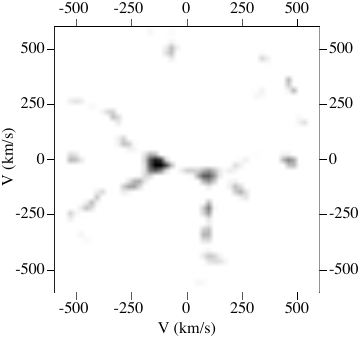}\\
\raisebox{2cm}{(b)} &
\includegraphics[width=2.5cm]{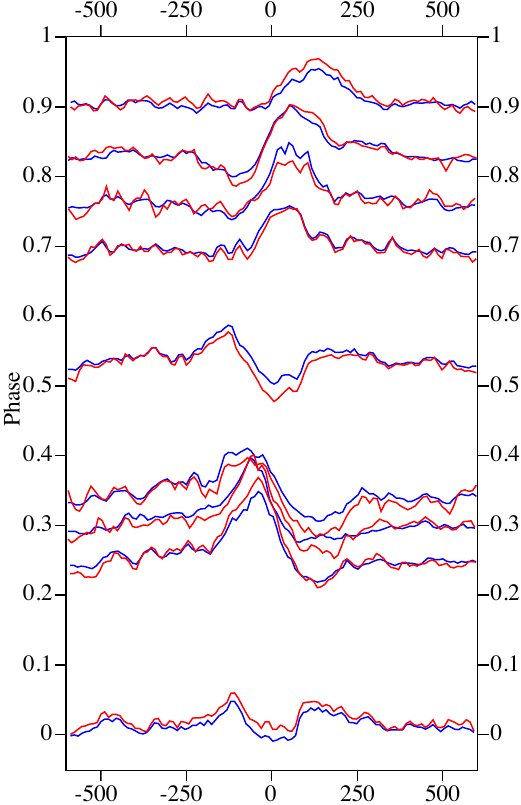} &
\includegraphics[width=2.5cm]{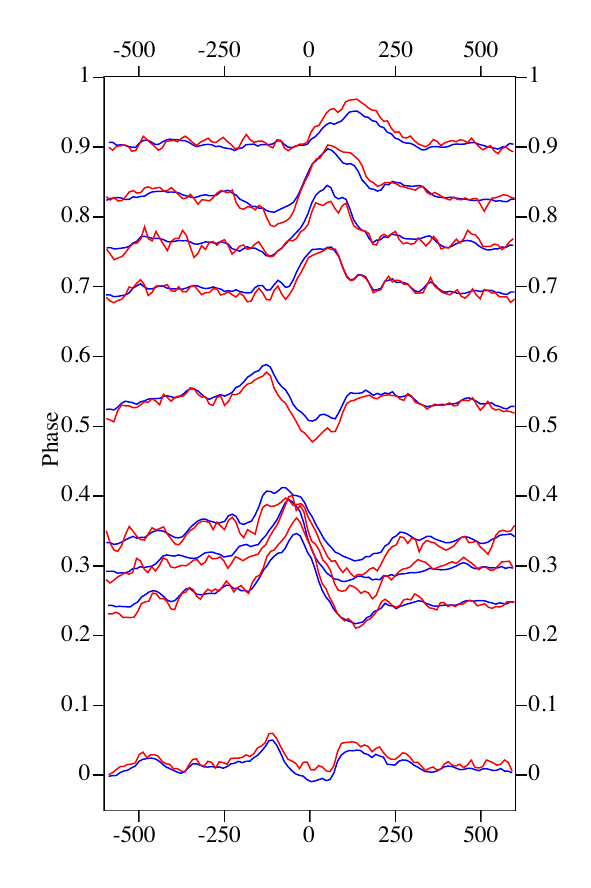} &
\includegraphics[width=2.5cm]{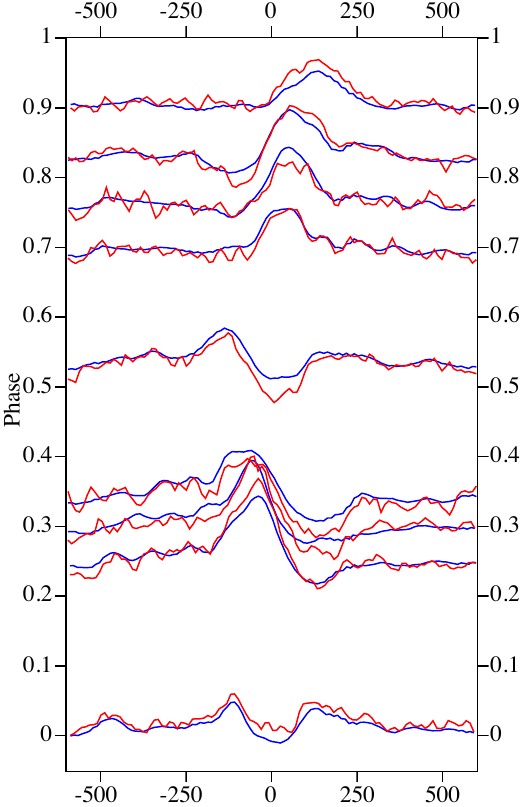} &
\includegraphics[width=2.5cm]{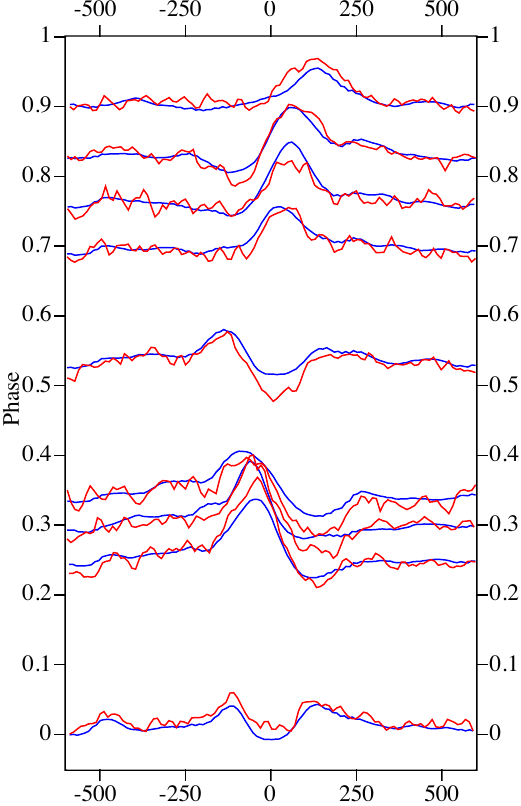} &
\includegraphics[width=2.5cm]{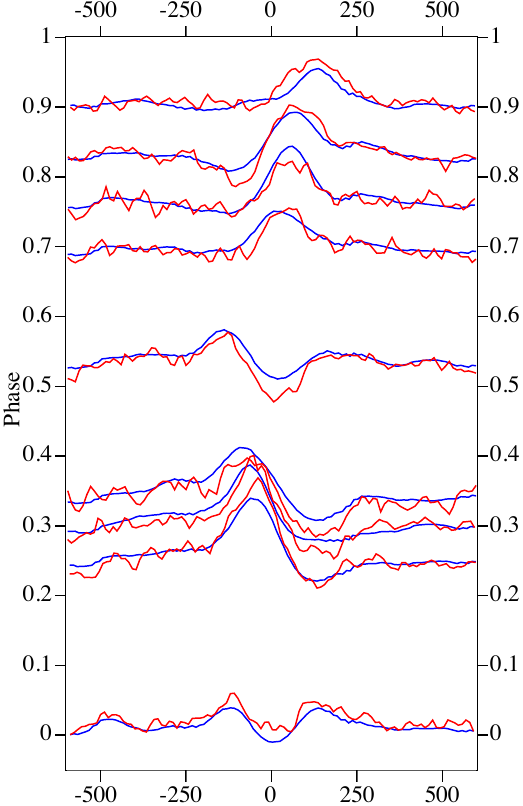}\\
\raisebox{1cm}{(c)} &
\includegraphics[width=2.5cm]{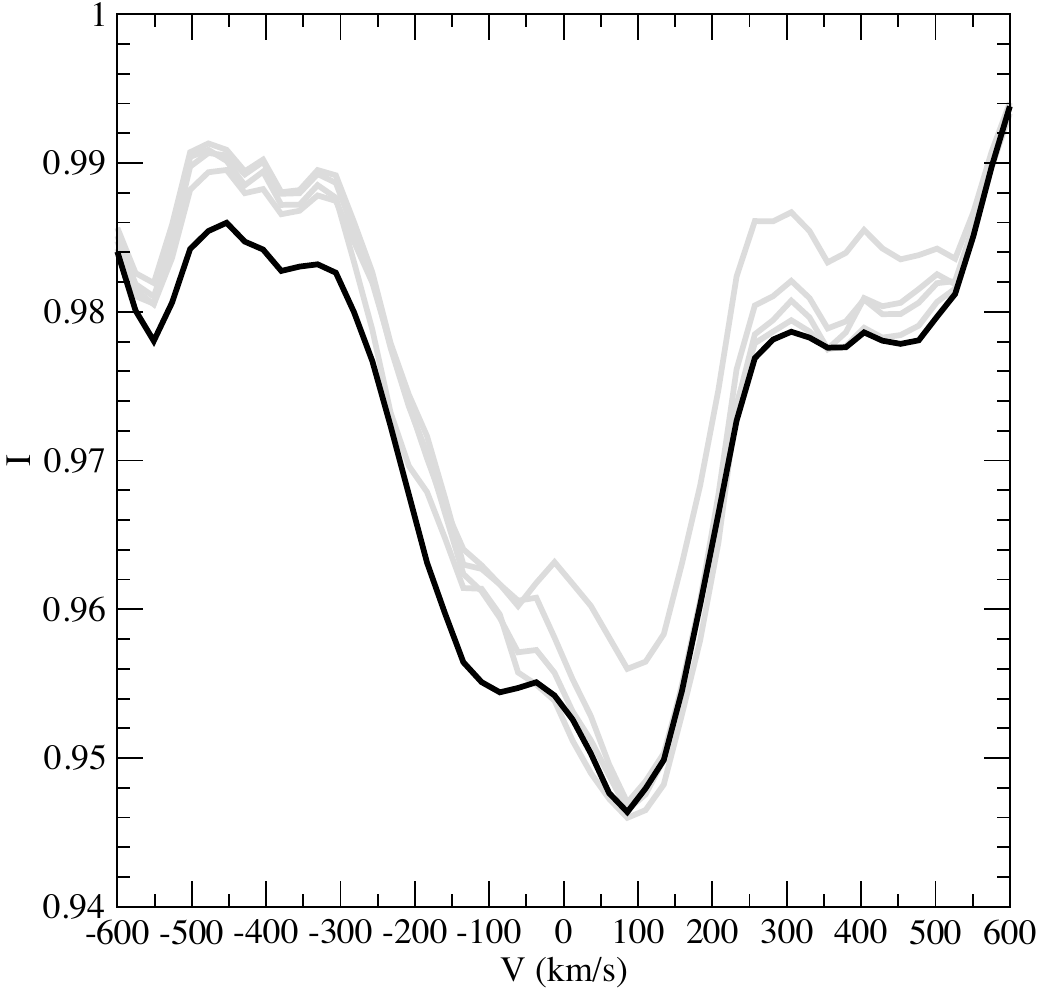} &
\includegraphics[width=2.5cm]{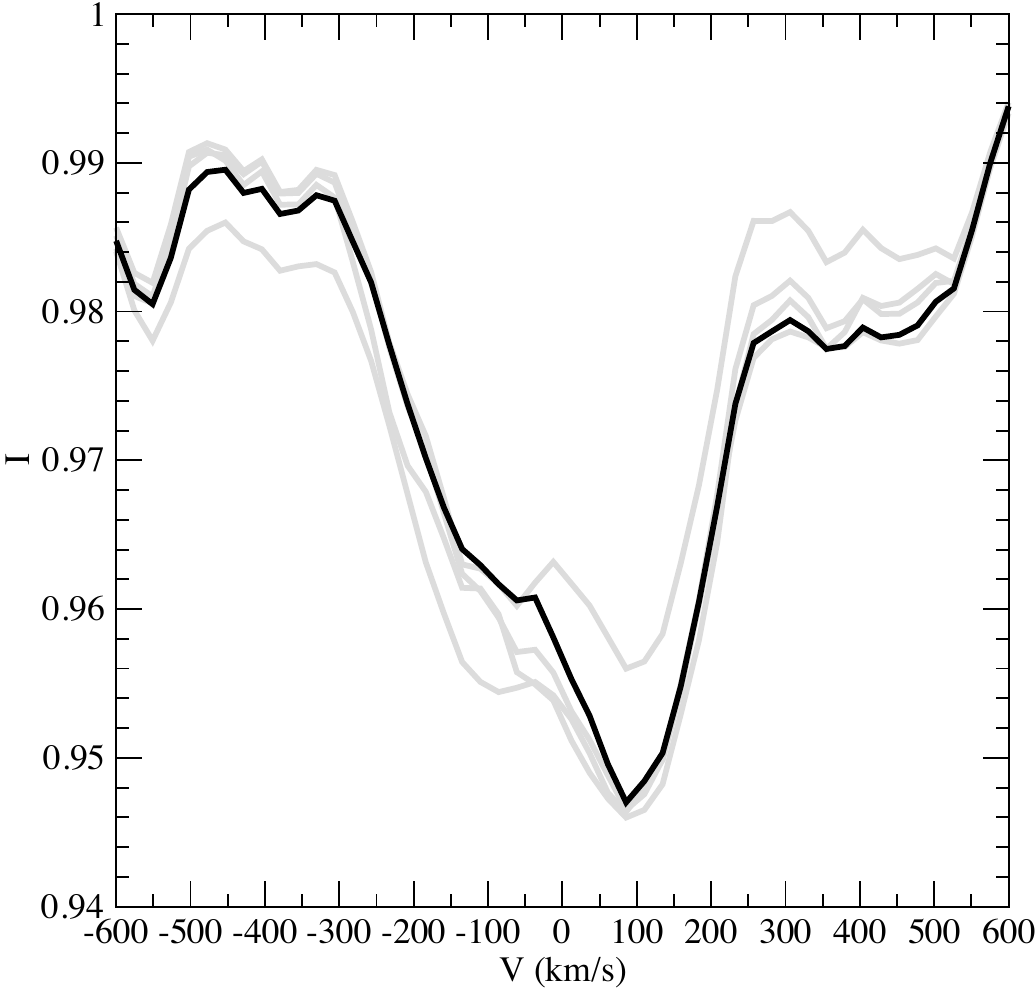} &
\includegraphics[width=2.5cm]{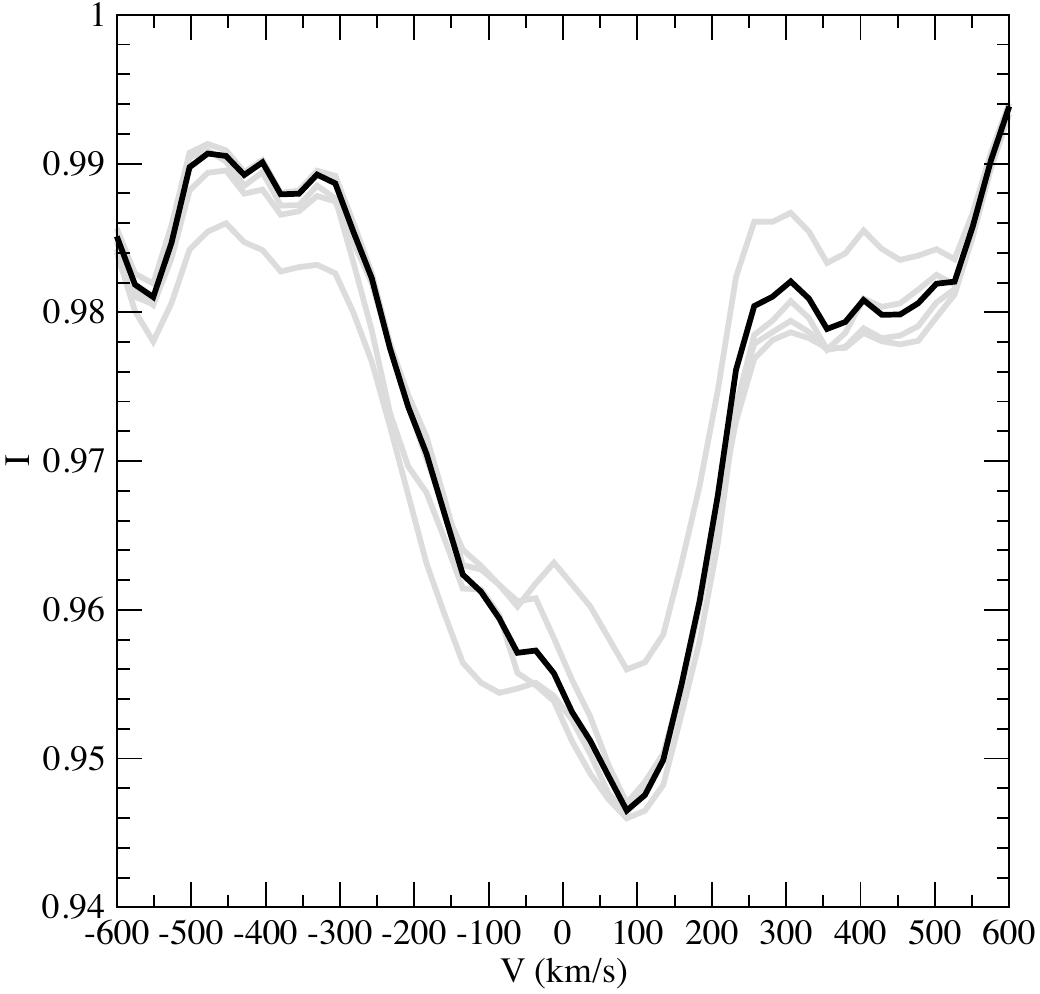} &
\includegraphics[width=2.5cm]{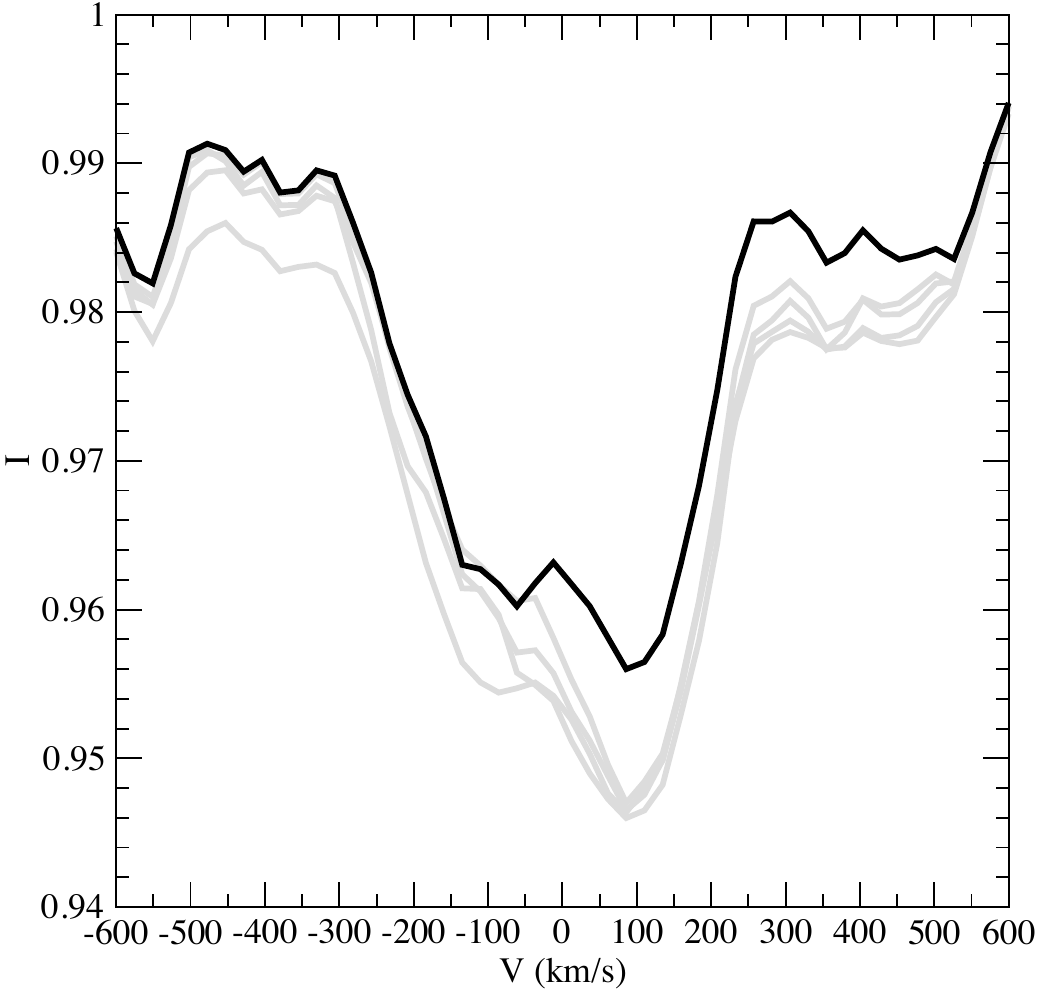} &
\includegraphics[width=2.5cm]{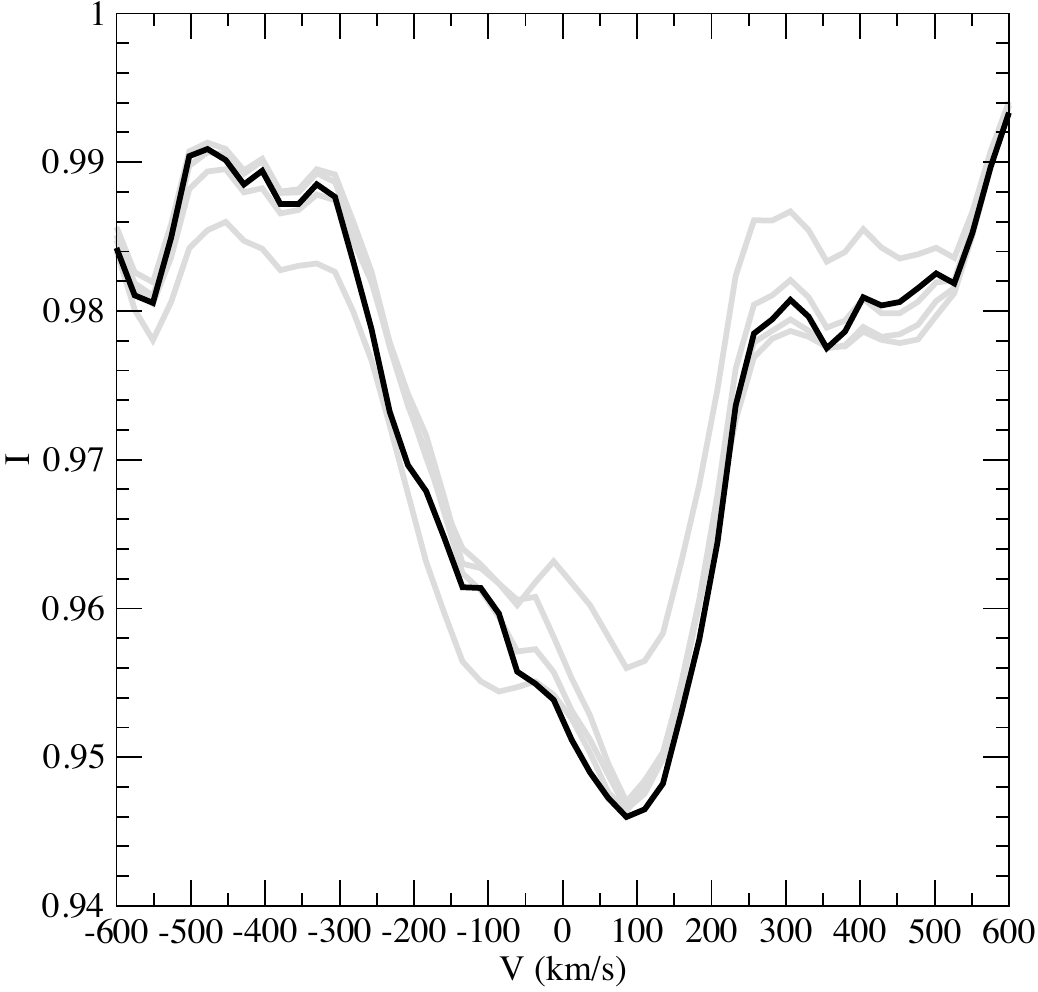} \\
\raisebox{1mm}{$\chi^2$} & 0.0139 & 0.0155 & 0.0166 & 0.0193 & 0.0195 \\
\end{tabular}
\caption{Reconstruction results for the He~II (4687~\AA) profiles of the star Cyg X-1 from~\cite{2005A&AT...24..383K} at different values of the FWHM of transfer function (parameter in the top row of the table). In row (b), observational profiles are shown in red, and synthetic profiles obtained from the tomogram are shown in blue. Row (c) shows the absorption profiles obtained by reconstructing the tomograms. In all cases, minimization was stopped after 500 steps.}
\label{CygX1}
\end{figure}

Tomograms reconstructed from the He~II (4687~\AA) profiles of the star Cyg X-1 from~\cite{2005A&AT...24..383K} at different FWHM values of the transfer function (parameter $W$ in the top row of the table) are shown Fig.~\ref{CygX1}. Also, in Fig.~\ref{CygX1} in line (b), the profiles themselves, observational (in red) and synthetic (in blue), are shown. The reconstruction was performed with the absorption enabled, and the absorption profiles are shown in the row (c), in this case, for clarity, all profiles are shown in gray on each graph, and the profile for this calculation is shown in black. A selection of a value $W$ requires knowledge, first of all, of the temperature of the gas in the flow and, sometimes, cannot be made unambiguously. To show how the reconstruction results depend on this parameter, tomograms reconstructed from the same data, but with different $W$ values are shown in Fig.~\ref{CygX1}. As can be seen from the figure, the increase in $W$ leads to a decrease in the number of details on the tomogram, while the details themselves become smaller. A side effect is also a reduction in background noise, as $W$ increases. This is because a wide transfer function cannot effectively fit narrow noise peaks. This, apparently, leads to an increase in the $\chi^2$ value with growth of $W$, while, visually, the coincidence of synthetic and observational profiles (row b) is approximately the same for all cases. The absorption profiles also turned out to be approximately the same for all $W$, although the case of $W=70\,\text{km/s}$ stands out somewhat. The profile in this case was a little shallower, meaning that less absorption was required to fit the data. This means that the tomogram image in this case gives a ``greater contribution'' to the fitting than in the others, and perhaps, this value is in better agreement with observations. It should be noted that the above analysis is valid only under the assumption that the system has a common envelope, in which absorption occurs. A discussion of the validity of this assumption is beyond the scope of the paper; the shown results are provided solely for the purpose of illustrating the operation of the algorithm.

\section{Analysis of Observational Tomograms Using Tomo-V Instruments}

\begin{figure}
\centering
  \includegraphics[height=3cm]{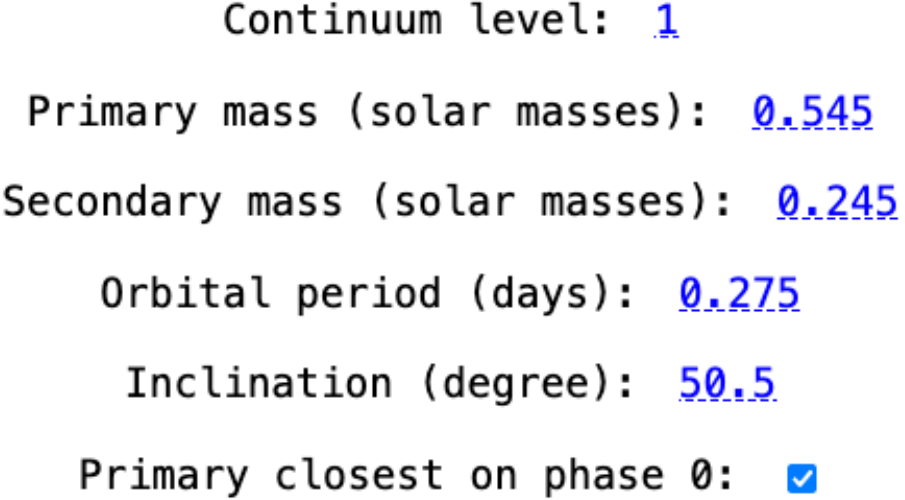} 
\caption{System parameters specified for SS Cyg.}
\label{params}
\end{figure}

Tomo-V has built-in tools for analyzing observational tomograms. Before using them, it is necessary to set the parameters of the system, for which the tomogram is being reconstructed: the masses of the primary and secondary components, the orbital period and the inclination of the orbit relative to the line of sight. It is also possible to specify the continuum level if it differs from unity in the profiles and to indicate whether the primary or secondary component is closer to the observer at phase zero. The parameters specified for the star SS Cyg are shown in Fig.~\ref{params}.

\begin{figure}[t]
\centering
\begin{tabular}{|c|c|}
\hline
\includegraphics[height=6cm]{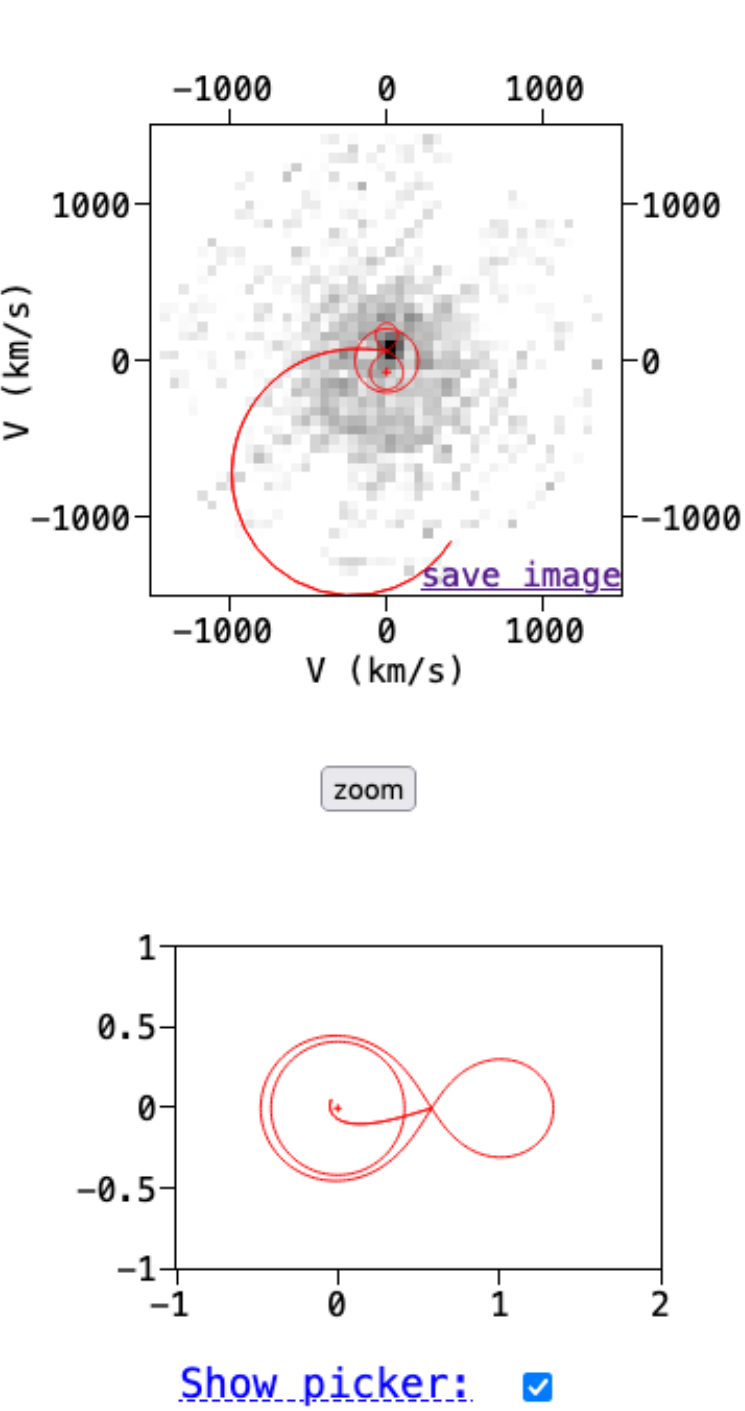} &
\includegraphics[height=6cm]{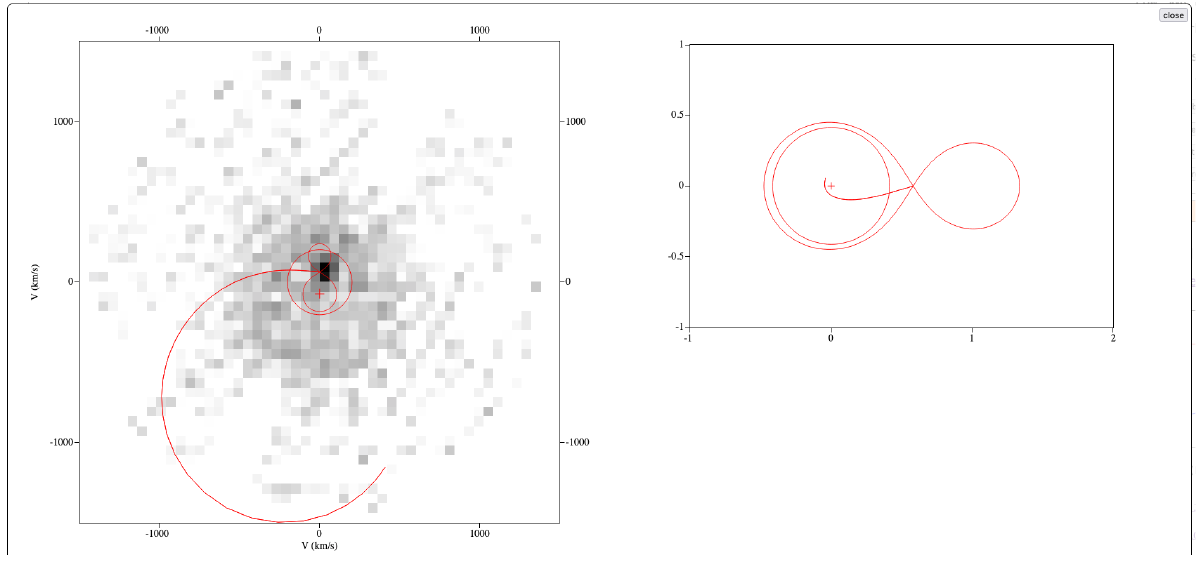} \\
\hline
\multicolumn{1}{c}{\rule{0pt}{5mm}(a)}&\multicolumn{1}{c}{(b)}
\end{tabular}

\caption{(a) Display of the ``Picker'' tool in the upper right part of the operating window, and (b) the result of pressing the [zoom] button. The display of the contours of the Roche lobes, the accretion disk of the main component, and the trajectory of the stream from the inner Lagrange point is included. For example, a tomogram for SS~Cyg is shown.}
\label{picker}
\end{figure}

To display the contours of the system’s elements on the tomogram: Roche lobes, accretion disks, as well as the trajectory of the stream from the inner Lagrange point $\mathrm{L_1}$, one can activate the options ``Show Primary Roche lobe'', ``Show Secondary Roche lobe'', ``Show Primary accretion disk'', ``Show Secondary accretion disk'', and ``Show Stream from $\mathrm{L_1}$'' in the right part of the operating window under the tomogram image. When displayed on a tomogram, these elements will be shown taking into account the system inclination. To show the same elements on the diagram in spatial coordinates, the ``Show picker'' checkbox should be activated. In Fig.~\ref{picker}, panel (a) shows the view of the right part of the window with the display options for elements and ``Picker'' enabled. Also, in Fig.~\ref{picker}, one can see the [zoom] button, when one clicks on it, the tomogram and ``Picker'' will be shown in an enlarged form, see panel (b) in the same figure. When elements are included, parameters can be specified for them. For example, for accretion disks, the eccentricity, semi-major axis, and angle between the semi-major axis of the disk and the system can be specified. By default, the eccentricity of accretion disks is set to zero, while the size of the semi-major axis of the disk is set to the radius of the last stable orbit in accordance with~\cite{1977ApJ...216..822P}:
$$
R_d=A\dfrac{0.6}{1+q}\qquad,
$$
where $A$ is the size of the major semi-axis of the system, and $q$ is the ratio of the component masses. The velocities in the disks are considered equal to the Keplerian ones. To calculate the velocities of the stream from the interior Lagrange point, a numerical solution of the three-body problem is performed.

For a stream from a point $\mathrm{L_1}$, the length and direction (towards the primary or secondary component) can be specified, and the display of the stream on the velocity field in the accretion disk can be enabled. Also, for each element, the color of the outline line and its thickness can be specified. When hovering the mouse cursor over any element in spatial coordinates (inside the ``Picker''), a mark as a circle with a crosshair, the center of which shows the local velocity. In this case, both the element itself and its analogue on the tomogram are clanging colours to blue.

\begin{figure}
\centering
\includegraphics[width=\linewidth]{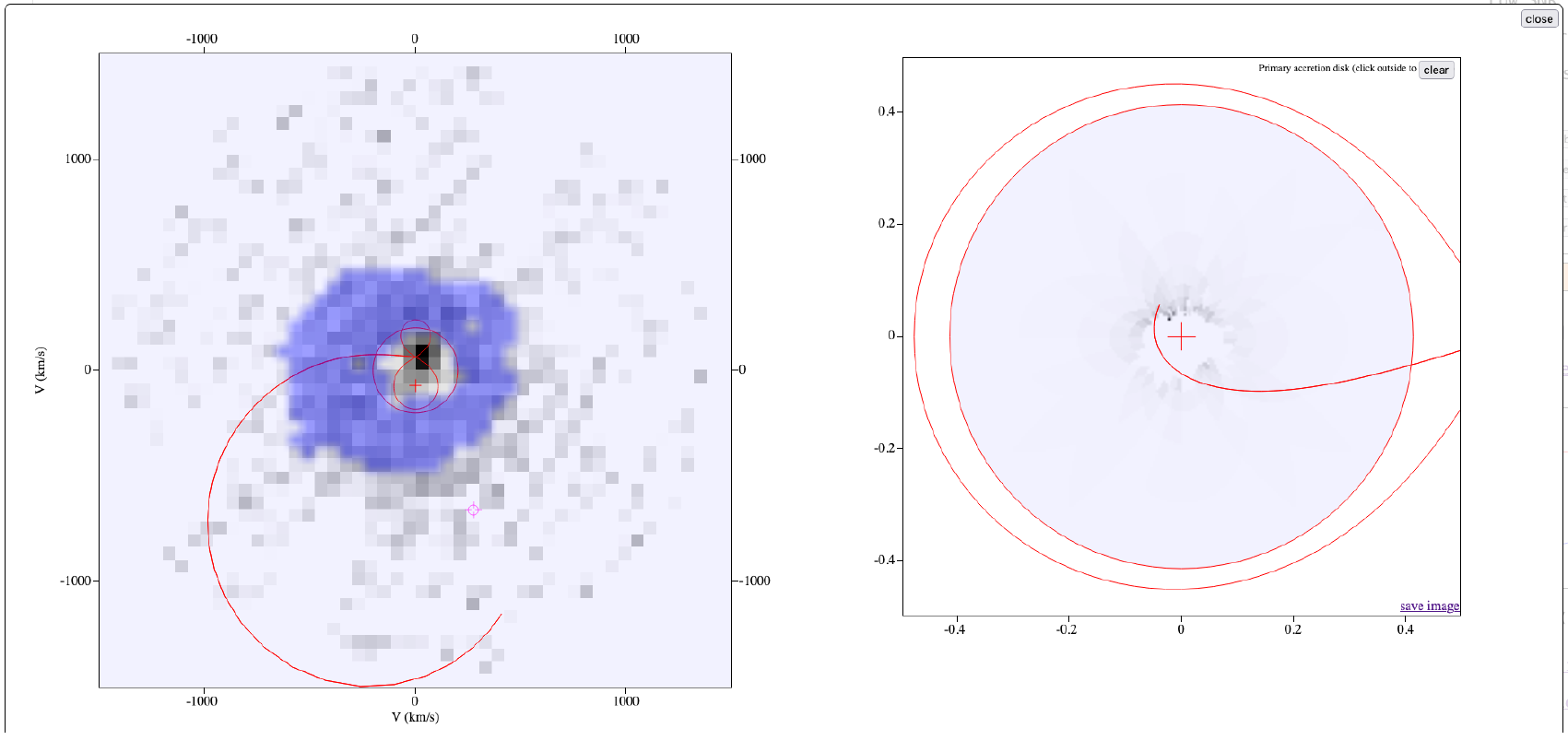}
\caption{Mapping brightness from a tomogram onto the accretion disk. For example, a tomogram for SS~Cyg is shown.}
\label{mapping}
\end{figure}

Another useful feature of Tomo-V is the ability to back-project brightness from the tomogram to system elements: accretion disks and Roche lobes. When clicking on any element in the ``Picker'', the selected element is enlarged, after which, by moving the mouse over the tomogram with the left button held down, one can ``paint'' the area of interest and its brightness will be displayed on the corresponding element in the ``Picker'' (see Fig.~\ref{mapping}). This takes into account the fact that each pixel on the tomogram can be mapped onto many pixels in spatial coordinates, and the brightness is uniformly ``smeared'' between them. This reduces the contribution of tomogram pixels located in the outer parts of the disk, since in this region, for each pixel of the tomogram, a larger surface area of the accretion disk, the brightness of which is summed up, is displayed. To clear the projection, click the [clear] button in the upper right corner of the ``Picker'', and to exit the display mode, click anywhere on the ``Picker'' outside the selected element.

\section{Conclusions}

The Tomo-V tool developed based on the algebraic reconstruction method and available on \url{https://tomo-v.inasan.ru} has shown good results in restoring tomograms based on both synthetic and real data. The method has coped well with blurred and noisy data and has been also capable of producing high-quality images from a small set of profiles. Furthermore, using this method, tomograms can be constructed from profiles contaminated with absorption lines, with the absorption profile being obtained as a by-product of tomogram reconstruction.

The tool made it possible to reduce the noise intensity in the initial data by using profile averaging. In this case, for averaging each profile, the opposite profile (obtained by interpolation), as well as a ``phantom'' profile obtained by interpolation between two adjacent profiles, can be taken. The interpolation is performed in Fourier space using the Central Projection Theorem. This method has its drawbacks, for example, it can lead to the appearance of artifacts near the ``corners'' of the image. Furthermore, this method cannot be used if the initial data contain absorption that makes the profiles asymmetric when the phase changes by $1/2$.

When restoring images from highly noisy data, the ``Low SNR'' mode, in which a certain cutoff threshold of $m<1$ is set, but as close to unity as possible, can be enabled. It is recommended to set $m>0.95$ to avoid significant artifacts. Selecting a value, it should be started with values around 0.99, gradually decreasing them until an image of satisfactory quality is obtained.

The used method makes it possible to construct clear images from blurred data. However, to obtain high-quality results, it is necessary to set correctly the full width at half maximum (FWHM) of the Gaussian transfer function. In the case, where this value is unknown, reconstruction should be performed with various FWHM values, paying attention to how the number and shape of features in the image changes, as well as the coincidence of the synthetic profiles with the observational ones, and then the best option should be selected. Indirectly, the selected value may be an indication of the temperature of the observed flow.

The performance of the program depends on the resolution of the tomogram (quadratically), on the FWHM value (quadratically), and on the total number of points in the profiles (linearly). Enabling absorption line fitting does not significantly affect the computing performance, except for the first step, when the initial values are calculated. Also, the time of the first step can be large, since all auxiliary tables that are used in further calculations are computed during it.

Almost every image on the screen can be saved to a file. To do this, it is necessary to hover the mouse cursor over it and click on the ``save image'' link that appears. A save dialog will open where the image settings can be adjusted. Saving can be done in either raster (PNG) or vector format by pressing the [print image] button. In the latter case, the browser’s standard print dialog, through which the image can be saved as a PDF file, will open.

The program state can be saved at any time by clicking the [save] button in the title bar of the operating window. The data are stored in the browser’s local storage. To share the data, it can exported to a file with the .tmv extension by clicking the [export] button. Such a file can be imported into the browser’s local storage by dragging it with the mouse to the area labeled ``drop .tmv file here'' in the left column. Attention: If there is insufficient disk space, the browser may clear the storage without warning. It may also be cleared when deleting cookies, so it is also recommended to save important data as tmv files. User data are not transmitted to the server in any way, are not stored there, and are not processed.

The ART method has proven its flexibility and effectiveness. In the future, the Tomo-V tool can be improved, for example, to reconstruct tomograms based on close lines that overlap each other. For this purpose, the transfer function can be specified not as a single Gaussian profile, but as a sum of shifted profiles according to the number of lines. In this way, separate tomograms can be obtained for each line. The same method can be used to restore three-dimensional tomograms; in this case, the line appears to be superimposed on itself, shifting due to the velocity component perpendicular to the orbital plane. However, to implement these modes, data, on which debugging and testing could be carried out, are needed.

Please note that Tomo-V is still under development, and some features not described in this paper may be added.

The author is open to collaboration. Please send all comments on the program, suggestions for development, and questions to the email address: \url{pasha@inasan.ru}

\section*{Acknowledgements}
The author expresses gratitude to everyone who helped in the development and debugging of the program by providing data for testing.

\section*{Funding}
The work on the development of Tomo-V was supported by the Russian Foundation for Basic Research, Russian-Bulgarian grant no. 20-52-18015/KP-06-Russia/2-2020 (Bulgarian National Science Fund).

\section*{Conflict of Interest}
The author of this work declares that he has no conflicts
of interest.

\bibliographystyle{az}
\bibliography{kaygorodov}

\end{document}